\documentclass[11pt,a4paper]{article}
\usepackage{jheppub} 
\pdfoutput=1
\usepackage{graphicx}
\usepackage{footmisc}
\usepackage{color}
\usepackage{amsmath} 
\usepackage{array,relsize,float}
\usepackage{url}
\usepackage[bitstream-charter]{mathdesign}
\usepackage{calc}
\usepackage{bm}
\usepackage{siunitx}
\usepackage{booktabs}
\usepackage{multirow}
\usepackage{subcaption}
\usepackage{tikz}
\usetikzlibrary{arrows.meta,positioning,fit,calc,backgrounds}
\usepackage{quantikz}

\renewcommand{\thefigure}{\arabic{figure}}

\usepackage{soul} 

\def\beq{\begin{equation}}
\def\eeq{\end{equation}}
\def\bea{\begin{eqnarray}}
\def\eea{\end{eqnarray}}
\def\beqn{\begin{eqnarray}} 
\def\eeqn{\end{eqnarray}}

\def\Eq#1{Eq.~(\ref{#1})}
\def\ln#1{\mathrm{log}\left(#1\right)}

\def\ket#1{|{#1}\rangle}

\newcommand{\valencia}{Instituto de F\'{\i}sica Corpuscular, Universitat de Val\`{e}ncia -- Consejo Superior de Investigaciones Cient\'{\i}ficas, Parc Cient\'{\i}fic, E-46980 Paterna, Valencia, Spain.}
\newcommand{\culiacan}{Facultad de Ciencias Naturales y Exactas, Universidad Autónoma de Sinaloa, Ciudad Universitaria, CP 80000 Culiacán, Mexico.}
\newcommand{\salamanca}{Departamento de Física Fundamental e IUFFyM, Universidad de Salamanca, 37008 Salamanca, Spain.}
\newcommand{\baires}{
Departamento de Física and IFIBA-CONICET, FCEyN, Universidad de Buenos Aires, Ciudad Universitaria, CP 1428 Buenos Aires, Argentina.}

\begin{document}
\title{A quantum representation of $\pi$ fragmentation functions through variational quantum circuits}
\author[a]{David~F. Renter\'{\i}a-Estrada,}
\author[b]{Roger~J. Hernández-Pinto,}
\author[a]{Germán Rodrigo,}
\author[c]{Rodolfo Sassot,}
\author[d]{and German~F.~R.~Sborlini} 
\affiliation[a]{\valencia}
\affiliation[b]{\culiacan}
\affiliation[c]{\baires}
\affiliation[d]{\salamanca}

\emailAdd{david.renteria@ific.uv.es}
\emailAdd{roger@uas.edu.mx}
\emailAdd{german.rodrigo@csic.es}
\emailAdd{sassot@df.uba.ar}
\emailAdd{german.sborlini@usal.es}

\preprint{MLQC4FC-2026-0002}

\abstract{We present a variational quantum-circuit model for fragmentation functions (FFs). Isospin and charge-conjugation symmetries are imposed to construct an independent six-flavor basis describing charged and neutral pion production, while physics-inspired Ansätze, including logarithmic feature maps and mass thresholds, encode the relevant kinematics. This quantum architecture substantially reduces the quantum circuit redundancies and improve optimization convergence. Using the DSS14 pion FF set as a benchmark, we first develop a one-dimensional variational representation (FF-VQR) in the momentum fraction at fixed energy scale, and show how entanglement between quark and gluon FFs yields a significant improvement, with accurate results already obtained using just two variational layers. A spectral analysis further demonstrates that the quantum model achieves high expressivity with a limited number of Fourier modes, supporting its use as a compact non-perturbative parametrization suitable for DGLAP evolution. We then extend the FF-VQR to two dimensions by incorporating the energy-scale dependence. By encoding all flavor channels within a single entangled quantum circuit, the quantum model provides a unified representation with higher accuracy than an independent encoding for each partonic species.}

\maketitle

\setcounter{page}{1}

\section{Introduction}
\label{sec:Introduction}
Fragmentation Functions (FFs) are fundamental non-perturbative quantities in Quantum Chromodynamics (QCD), describing the probability that a parton produced in a short-distance interaction hadronizes into an observed hadron carrying a fraction of the parent parton's momentum \cite{Field:1976ve,Collins:1989gx,Collins:2023cuo}. Together with Parton Distribution Functions (PDFs), FFs constitute an essential ingredient of the QCD factorization theorem, enabling the separation of perturbatively calculable hard-scattering cross sections from the long-distance dynamics associated with hadron formation \cite{Collins:1989gx}.

The determination of FFs plays a central role in modern particle physics. They are required for the theoretical description of a broad class of observables involving identified hadrons in the final state, ranging from electron-positron annihilation to deep-inelastic scattering and hadron-hadron collisions \cite{ALEPH:1994cbg,DELPHI:1998cgx,OPAL:1994zan,SLD:1998coh,TASSO:1988jma,HERMES:2012uyd,Makke:2013bya,PHENIX:2003fvg,BRAHMS:2007tyt,STAR:2006dgg}. Precise knowledge of FFs is also crucial for interpreting measurements performed at BNL's Relativistic Heavy Ion Collider (RHIC) and the CERN's Large Hadron Collider (LHC), where identified pions, kaons, and other hadrons are routinely employed as probes of perturbative and non-perturbative QCD dynamics \cite{ALICE:2012wos,ALICE:2017nce,ALICE:2021est,Borsa:2021ran}.

Historically, FFs have been extracted through global QCD analyses combining experimental information from several processes and energy scales. Significant progress has been achieved through the DSS~\cite{deFlorian:2007ekg} framework and its subsequent extensions, leading to increasingly precise determinations of pion, kaon, proton, and charged-hadron FFs~\cite{deFlorian:2014xna,deFlorian:2017lwf,Borsa:2022vvp}. More recently, neural-network approaches have also been introduced to reduce parametrization biases and improve uncertainty estimation in  FF analyses~\cite{Bertone:2018ecm,Moffat:2021dji,Soleymaninia:2020bsq,Soleymaninia:2022qjf,Soleymaninia:2024jam,Soleymaninia:2026xjq,Galvez-Viruet:2026jgx}.

Conventional FF determinations still relies on selecting suitable functional forms at  a reference energy scale and optimizing their parameters through fitting procedures. The flexibility of the chosen Ansatz directly affects the quality of the final description and may introduce non-negligible model dependence. Similar challenges motivated the development of neural-network parametrizations in the PDF community, where machine-learning techniques have demonstrated the capability to provide highly flexible and unbiased functional representations~\cite{Ball:2008by,Ball:2010de,Carrazza:2019mzf,NNPDF:2019vjt}.

The rapid development of Quantum Machine Learning (QML) opens the possibility of extending these ideas to quantum-computing architectures. In particular, Variational Quantum Circuits (VQCs) have emerged as powerful quantum models capable of acting as universal function approximators through the combination of data encoding, trainable quantum operations, and quantum interference effects \cite{Perez-Salinas:2019pjx,Perez-Salinas:2020nem,Schuld:2020enb,Cerezo2021VariationalQuantumAlgorithms}. These architectures naturally generate highly non-linear feature maps in Hilbert space and offer a compact representation of complex functions through a relatively small number of trainable parameters.

Several studies have recently explored the application of QML techniques to problems in High-Energy Physics (HEP) and QCD phenomenology. In particular, VQCs have been proposed for the determination of proton structure functions and effective partonic distributions \cite{Perez-Salinas:2020nem,Ochoa-Oregon:2024zgm}. Furthermore, the first investigations of parton-fragmentation dynamics on quantum computers have recently appeared in the literature, highlighting the potential of quantum algorithms for the study of hadronization phenomena~\cite{deLejarza:2025upd,Li:2024nod}.

In this work, we investigate the capability of VQCs to represent charged-pion FFs. Rather than performing a direct fit to experimental measurements, our goal is to study the expressive power of quantum models by learning established  FF parametrizations. Using the DSS14 pion FFs~\cite{deFlorian:2014xna} as a reference set, we construct quantum models capable of accurately reproducing both the one-dimensional dependence on the momentum fraction $z$ at fixed energy scale $Q_0$, and the full two-dimensional dependence on $(z,Q^2)$.

The proposed VQC framework incorporates physics-inspired structures directly into the quantum architecture. Heavy-flavor thresholds, gluon-enhanced encodings, and flavor-depen\-dent entanglement patterns are embedded into the circuit design, allowing the quantum model to exploit prior knowledge from perturbative QCD evolution while retaining the flexibility of variational quantum learning. In this way, the quantum circuit does not learn the FFs from the ground up, but rather develops a compact quantum representation of the non-perturbative structures governing hadronization.

An additional objective of this work is to analyze the learned quantum representations from a spectral perspective. It has recently been shown that VQCs can be interpreted as truncated Fourier expansions whose accessible frequency spectrum is determined by the encoding strategy and quantum circuit architecture \cite{Schuld:2020enb,Wiedmann:2024fourier,Atchade-Adelomou:2023mjf}. By studying the Fourier decomposition of the trained quantum observables, we investigate the spectral content required to describe FFs and identify the origin of residual discrepancies appearing in strongly suppressed kinematic regions.

The paper is organized as follows. First, we review the classical methods to extract FFs from experiments in Sec. \ref{sec:classical_determination}. Then, we introduce a Variational Quantum Representation of FFs (FF-VQR) at a fixed energy scale, and benchmark its performance against the DSS14 parametrization in Sec. \ref{sec:vqc_q0}. We investigate the spectral structure generated by the VQCs through a Fourier analysis of the learned observables in Sec. \ref{sec:SpectralVQC}, in order to both justify the expressivity of the models and provide a compact representation. After that, we extend the VQC framework to the full two-dimensional $(z,Q^2)$ kinematic domain, incorporating heavy-flavor thresholds, in Sec. \ref{sec:sevqc_interpolator}. Finally, in Sec. \ref{sec:Conclusions}, we present the conclusions and discuss future research directions.

\section{Classical determination of fragmentation functions}
\label{sec:classical_determination}
Fragmentation functions are non-perturbative quantities that characterize the hadronization of quarks and gluons into identified hadrons. They encode the probability for a parton $i$ to hadronize into a hadron $h$ carrying a fraction $z$ of the parent parton's momentum at a factorization energy scale $Q$~\cite{Field:1976ve,Collins:1989gx}. Within the framework of QCD factorization, their functional form must be extracted from experimental measurements corresponding to several classes of hard-scattering processes. The cleanest environment is provided by single-inclusive electron--positron annihilation (SIA),
\begin{equation}
e^+e^- \rightarrow h + X,
\end{equation}
where an identified hadron $h$ is observed in the final state while all remaining particles are left unobserved. Experimentally, the relevant observable is the normalized differential cross section as a function of the hadron energy fraction
\begin{equation}
z=\frac{2E_h}{\sqrt{s}},
\end{equation}
which measures the fraction of the available center-of-mass energy carried by the observed hadron. Since no hadrons appear in the initial state, SIA provides the most direct probe of fragmentation dynamics. The normalized SIA cross section can be expressed as a convolution of the FFs,
\begin{equation}
\frac{1}{\sigma_{\rm tot}}\frac{d\sigma^h}{dz}
=
\sum_i
C_i(z,\alpha_s(Q^2))
\otimes
D_i^{h}(z,Q^2),
\end{equation}
where $C_i$ are perturbatively calculable coefficient functions, $D_i^h$ are the FFs, and $\otimes$ denotes the Mellin convolution. Measurements from LEP, SLC, Belle, and BaBar constitute the primary source of information on the flavor-singlet sector~\cite{deFlorian:2007ekg,deFlorian:2014xna,deFlorian:2017lwf}. 

Additional flavor sensitivity is obtained from semi-inclusive deep-inelastic scattering (SIDIS),
\begin{equation}
\ell + N \rightarrow \ell + h^{\pm} + X,
\end{equation}
where a lepton scatters off a nucleon target and a hadron is detected in coincidence with the scattered lepton. In contrast to SIA, SIDIS depends simultaneously on the partonic structure of the nucleon and on the fragmentation process. The relevant kinematic variables are
\begin{equation}
x =
\frac{Q^2}{2p\cdot q},
\qquad
z_h =
\frac{p_h\cdot p}{q\cdot p},
\qquad
y =
\frac{Q^2}{xs},
\end{equation}
where $x$ denotes the Bjorken scaling variable, $z_h$ is the fraction of the virtual-photon energy carried by the detected hadron, and $y$ is the inelasticity variable. The SIDIS differential cross section can be written in terms of two structure functions, $F_1^h$ and $F_L^h$, according to
\begin{equation}
\frac{d\sigma^h}{dx\,dy\,dz_h}
=
\frac{2\pi\alpha^2}{Q^2}
\left[
\frac{1+(1-y)^2}{y}
\,2F_1^h(x,z_h,Q^2)
+
\frac{2(1-y)}{y}
F_L^h(x,z_h,Q^2)
\right].
\end{equation}
At next-to-leading order (NLO), the transverse and longitudinal structure functions are given by
\begin{align}
F_1^h
&=
\frac{1}{2}
\sum_{q,\bar q}
e_q^2
\left\{
f_q^N \otimes D_q^h
+
\frac{\alpha_s}{2\pi}
\left[
f_q^N\otimes C_{qq}^{1}\otimes D_q^h
+
f_q^N\otimes C_{gq}^{1}\otimes D_g^h
+
f_g^N\otimes C_{qg}^{1}\otimes D_q^h
\right]
\right\}, \\
%
F_L^h
&=
\frac{\alpha_s}{2\pi}
\sum_{q,\bar q}
e_q^2
\left[
f_q^N\otimes C_{qq}^{L}\otimes D_q^h
+
f_q^N\otimes C_{gq}^{L}\otimes D_g^h
+
f_g^N\otimes C_{qg}^{L}\otimes D_q^h
\right].
\end{align}
Here, $f_q^N$ and $f_g^N$ denote the quark and gluon PDF of the nucleon, while $C_{ij}^{1,L}$ are perturbatively calculable coefficient functions. 

Further constraints on FFs are obtained from single-inclusive hadron production in hadronic collisions,
\begin{equation}
p_1(P_A)+p_2(P_B)\rightarrow h(P_h)+X,
\end{equation}
where a hadron $h$ is detected in the final state with transverse
momentum $p_T$ and rapidity $\eta$. In contrast to SIA and SIDIS,
hadronic collisions receive sizable contributions from gluon-initiated
subprocesses and therefore provide direct sensitivity to the gluon
FF. Within the QCD factorization framework, the invariant cross section can
be expressed as
\begin{equation}
E_h\frac{d^3\sigma^{h}}{dp_h^3}
=
\sum_{a,b,c}
f_a^{h_1}(x_a,\mu_I^2)
\otimes
f_b^{h_2}(x_b,\mu_I^2)
\otimes
d\hat{\sigma}_{ab\rightarrow cX}
(x_a,x_b,z,
\mu_R,\mu_I,\mu_{F})
\otimes
D_c^h(z,\mu_{FF}^2),
\end{equation}
where $f_a$ and $f_b$ denote the PDFs of
the incoming protons, $d\hat{\sigma}_{ab\rightarrow cX}$ represents
the perturbatively calculable partonic hard-scattering cross section,
and $D_c^h$ is the fragmentation function describing the transition
of the final-state parton $c$ into the observed hadron $h$.
The $\mu_R$, $\mu_I$, and $\mu_{F}$ denote the renormalization, initial-state factorization, and final-state factorization scales, respectively. Measurements of charged and identified hadron production at RHIC and the LHC provide important constraints on the gluon FF, particularly at intermediate and large values of $z$, thereby complementing the information obtained from SIA and SIDIS
data~\cite{PHENIX:2003fvg,BRAHMS:2007tyt,STAR:2006dgg}. 

Although FFs are intrinsically non-perturbative objects, their dependence on the factorization scale is predicted by perturbative QCD through the timelike Dokshitzer--Gribov--Lipatov--Altarelli--Parisi (DGLAP) evolution equations~\cite{Dokshitzer:1977sg, Gribov:1972ri,Altarelli:1977zs},
\begin{equation}
    \frac{d}{d\operatorname{ln}Q^2}{\vec D}^h_i(z,Q^2)=\sum_j\left(P_{ij}\otimes {\vec D}_j\right)(z,Q^2)\,, 
\end{equation}
\begin{equation}
    \frac{d}{d\operatorname{ln}Q^2}
    \begin{pmatrix}
        D_\Sigma^{H} \\
        D_g^{H}
    \end{pmatrix} =  \begin{pmatrix}
                        P_{qq}^T & 2n_fP_{gq}^T \\
                        \frac{1}{2n_f}P_{qg}^T & P_{gg}^T
                    \end{pmatrix}
                    \otimes
                    \begin{pmatrix}
                            D_\Sigma^{H} \\
                            D_g^{H}
                    \end{pmatrix}
                    \,,
                    \end{equation}
where $P_{ij}^{T}(z)$ denote the timelike splitting functions calculable in perturbative QCD. The singlet FF is defined as
\begin{equation}
D_\Sigma^{h}
=
\sum_q
\left(
D_q^{h}
+
D_{\bar q}^{h}
\right).
\end{equation}
These evolution equations introduce strong quark--gluon mixing and determine the scale dependence of FFs over a broad kinematic range. Once an initial parametrization is specified at a reference scale $Q_0^2$, the DGLAP equations allow the FFs to be evolved to any experimentally relevant scale. A fundamental consistency condition is provided by momentum conservation during the hadronization process, which leads to the momentum sum rule,
\begin{equation}
\sum_h
\int_0^1 dz \,
z\,D_i^h(z,Q^2)
=1.
\end{equation}
Modern global analyses combine SIA, SIDIS, and hadron-collision measurements within this factorization framework to determine FFs at an reference scale and subsequently evolve them using the timelike DGLAP equations~\cite{deFlorian:2014xna,Borsa:2022vvp}. The resulting distributions are typically provided in the form of interpolation grids. In the present work, these interpolated FFs are employed as target distributions for the construction of a compact quantum representation based on VQCs.

The charged-pion FF requires specifying the flavor basis together with the charge-conjugation and isospin relations. In this work, the FFs associated to positively charged pions are taken as independent distributions, while the corresponding negatively charged ones are obtained through charge conjugation. The valence structures of the charged pions are $\ket{\pi^+} = \ket{u\bar d}$ and $\ket{\pi^-}=\ket{\bar ud}$, so that charge conjugation implies
\begin{eqnarray}
D_q^{\pi^-}(z,Q^2) &=& D_{\bar q}^{\pi^+}(z,Q^2),
\label{eq:pi-q}\\
D_{\bar q}^{\pi^-}(z,Q^2) &=& D_q^{\pi^+}(z,Q^2),
\label{eq:pi-qbar}\\
D_g^{\pi^-}(z,Q^2) &=& D_g^{\pi^+}(z,Q^2).
\label{eq:pi-g}
\end{eqnarray}
Assuming isospin symmetry, the neutral-pion FFs are obtained from
\begin{equation}
D_i^{\pi^0}(z,Q^2)
=
\frac12
\left[
D_i^{\pi^+}(z,Q^2)
+
D_i^{\pi^-}(z,Q^2)
\right].
\end{equation}
Although the DSS14 interpolator provides the complete flavor basis, several FFs are related by charge-conjugation symmetry. Consequently, it is not necessary to assign independent variational parameters to each flavor channel. Instead, the FF-VQR parametrizes only one representative of every degenerate pair, reducing the number of variational degrees of freedom while preserving the complete flavor structure of the DSS14 parametrization. The independent FFs used by the FF-VQR are chosen as
\begin{equation}
\left\{
D_u^{\pi^+},
D_d^{\pi^+},
D_s^{\pi^+},
D_c^{\pi^+},
D_b^{\pi^+},
D_g^{\pi^+}
\right\},
\end{equation}
while the remaining FFs are reconstructed through
\begin{eqnarray}
D_{\bar d}^{\pi^+} &=& D_u^{\pi^+},\\
D_{\bar u}^{\pi^+} &=& D_d^{\pi^+},\\
D_{\bar s}^{\pi^+} &=& D_s^{\pi^+},\\
D_{\bar c}^{\pi^+} &=& D_c^{\pi^+},\\
D_{\bar b}^{\pi^+} &=& D_b^{\pi^+},
\end{eqnarray}
and the corresponding $\pi^-$ FFs are then obtained from Eqs.~(\ref{eq:pi-q})-(\ref{eq:pi-g}) through charge conjugation.

\section{Variational quantum representation at a fixed energy scale}
\label{sec:vqc_q0}
The objective of this section is to investigate how FFs obtained from a state-of-the-art global analysis can be efficiently represented by VQCs. To this end, the interpolation grids provided by the DSS14 analysis~\cite{deFlorian:2014xna} are employed as target distributions. The quantum model is therefore trained to reproduce the functional dependence of the FFs, replacing the classical interpolation by a compact quantum representation. As a first step, we consider FFs at a fixed energy scale,
\begin{equation}
Q^2 = Q_0^2,
\end{equation}
such that the problem reduces to learning one-dimensional functions of the hadron momentum fraction. For each partonic flavor,
\begin{equation}
i=\{u,d,s,c,b,g\},
\label{eq:FlavorLabel}
\end{equation}
the target distribution is given by the corresponding DSS14 interpolation $D_i^h(z,Q_0^2)$. Since FFs span several orders of magnitude and vanish in the kinematic limit $z\rightarrow1$, the quantum model is trained using
\begin{equation}
zD_i^h(z,Q_0^2),
\end{equation}
which exhibits a smoother behavior throughout the physical region and reduces the numerical impact of endpoint singularities.

The variational quantum model employs one qubit per partonic flavor. For a given value of~$z$, the input variable is encoded through a nonlinear feature map motivated by the logarithmic behavior typically observed in FFs,
\begin{equation}
\bm\omega \cdot \bm x(z)
=
\omega_{1i} z
+
\omega_{2i} \log z
+
\omega_{3i} \ln{1-z}
+
\omega_{4i},
\label{eq:Representation}
\end{equation}
where $\omega_{1i}$, $\omega_{2i}$, $\omega_{3i}$, and $\omega_{4i}$ are trainable parameters. Each qubit is initialized in an equal-superposition state through a Hadamard gate and subsequently evolved according to
\begin{equation}
\label{eq:state1}
|\psi_i\rangle
=
{\rm Rot}(\alpha_i,\beta_i,\gamma_i)
R_Y(\bm\omega\cdot\bm x(z))
H|0\rangle ,
\end{equation}
where the trainable operator is defined as ${\rm Rot}(\alpha_i,\beta_i,\gamma_i)\equiv R_Z(\alpha_i)R_Y(\beta_i)R_Z(\gamma_i)$. A quantum circuit constructed exclusively from single-qubit operations would represent each flavor channel independently. To increase the expressive power of the Ansatz, successive layers of local rotations are supplemented by controlled-$R_Y$ operations connecting the gluon qubit to the quark-flavor qubits,
\begin{equation}
U_{\rm ent}
=
{\rm CRY}_{u\rightarrow g}(\phi_{ug})
\prod_q {\rm CRY}_{g\rightarrow q}(\phi_{gq})
\label{eq:Uent}
\end{equation}
being $\phi_q$ trainable parameters. Here, the gluon qubit acts as the control and the quark qubits as targets, while an additional connection uses the up-quark qubit as the control and the gluon qubit as the target. The controlled-$R_Y$ gates generate conditional phase correlations between the gluon and quark channels. For a control qubit prepared in a superposition state, the action of a controlled-$R_Y$ gate produces states of the form
\begin{equation}
\label{eq:state2}
\ket{\Psi}
= {\rm CRY}(\phi)(A|0\rangle+B|1\rangle))\otimes |\psi\rangle =
\left( A|0\rangle \otimes I
+
B|1\rangle \otimes R_Y(\phi) \right)|\psi\rangle ,
\end{equation}
with $A$ and $B$ complex amplitudes. As a consequence, the quantum state can no longer be described as a simple product of independent flavor contributions. The information encoded in one channel becomes distributed across several qubits, allowing the quantum circuit to build more complex representations than those accessible through independent single-qubit models. From a variational perspective, the entangling layers enlarge the accessible manifold of correlated quantum states and effectively allow the parameters associated with different flavor channels to contribute collectively to the final prediction. This generates a richer set of nonlinear combinations and correlations, enhancing the representational capacity of the quantum model while requiring only a modest increase in the number of trainable parameters. The choice of the gluon qubit as the central node is motivated by the prominent role played by the gluon in the timelike evolution equations, where gluon and quark channels are strongly coupled through QCD evolution kernels. Since we are fitting $\pi^+$ FFs, which depend strongly on the up-quark distribution, we assign a privileged role to this flavor by introducing the ${\rm CRY}_{u\to g}$ gate in Eq. (\ref{eq:Uent}). Although the present study is performed at a fixed scale and does not explicitly solve the DGLAP equations, this interconnected pattern provides a physically motivated structure for the variational Ansatz. 

The prediction associated with each flavor channel is obtained from the expectation value of the Pauli-$Z$ operator,
\begin{equation}
\mathcal O_i(z)
=
\langle Z_i\rangle ,
\label{eq:Observable}
\end{equation}
which is mapped onto the interval $[0,1]$ according to
\begin{equation}
z\,\widetilde D_i(z)
=
\frac{1+\mathcal O_i(z)}{2}.
\end{equation}
To match the range of the quantum prediction, the DSS14 targets are
independently normalized for each flavor channel according to
\begin{equation}
z\,\widetilde D_i^{\rm DSS14}(z,Q_0^2)
=
\frac{
z\,D_i^{\rm DSS14}(z,Q_0^2)
}{
\displaystyle
\max_{z}
\left[
z\,D_i^{\rm DSS14}(z,Q_0^2)
\right]
},
\end{equation}
such that
$z\widetilde D_i^{\rm DSS14}(z,Q_0^2)\in[0,1]$.
The trainable parameters are determined by minimizing the mean-squared error
(MSE) between the quantum prediction and the normalized DSS14 interpolation
grid, i.e.
\begin{equation}
\mathcal L
=
\frac{1}{N_zN_f}
\sum_{k=1}^{N_z}
\sum_{i=1}^{N_f}
\left[
z\,\widetilde D_i(z_k)
-
z\,\widetilde D_i^{\rm DSS14}(z_k,Q_0^2)
\right]^2.
\end{equation}
where $N_z$ denotes the number of sampling points and $N_f$ the number of active flavors. Unlike conventional global analyses, no experimental information enters the optimization procedure. The VQC is trained exclusively on the interpolated FFs, allowing us to quantify how efficiently a quantum model can compress and reproduce the non-perturbative structures encoded in modern FF determinations. The fixed-scale study serves as a benchmark for the more challenging two-dimensional problem discussed in Section~\ref{sec:sevqc_interpolator}, where the quantum model is required to reproduce the simultaneous dependence on both $z$ and $Q^2$. 

\begin{table}[t]
\centering
\begin{tabular}{c c c c c}
\hline
Layers & Trainable parameters
& $\mathcal{L}_{\rm MSE}(Q_1^2)$
& $\mathcal{L}_{\rm MSE}(Q_2^2)$
& $\mathcal{L}_{\rm MSE}(Q_3^2)$ \\
\hline
1 & 48 & $7.0\times 10^{-4}$ & $4.4\times 10^{-5}$ & $4.2\times 10^{-5}$ \\
2 & 96 & $3.0\times 10^{-4}$ & $3.2\times 10^{-6}$ & $5.8\times 10^{-6}$ \\
3 & 144 & $1.6\times 10^{-4}$ & $3.5\times 10^{-6}$ & $3.5\times 10^{-6}$ \\
\hline
\end{tabular}
\caption{
Reconstruction performance of the FF-VQR at a fixed scale: $Q_1^2=10$ GeV$^2$, $Q_2^2=100$ GeV$^2$ and $Q_3^2=1000$ GeV$^2$. The model was trained using a learning rate of $1\times10^{-2}$. The table reports the number of variational layers, the total number of trainable parameters, and the final normalized MSE.}
\label{tab:vqr_depth_scan}
\end{table}
\begin{figure}
    \centering
\includegraphics[width=0.99\linewidth]{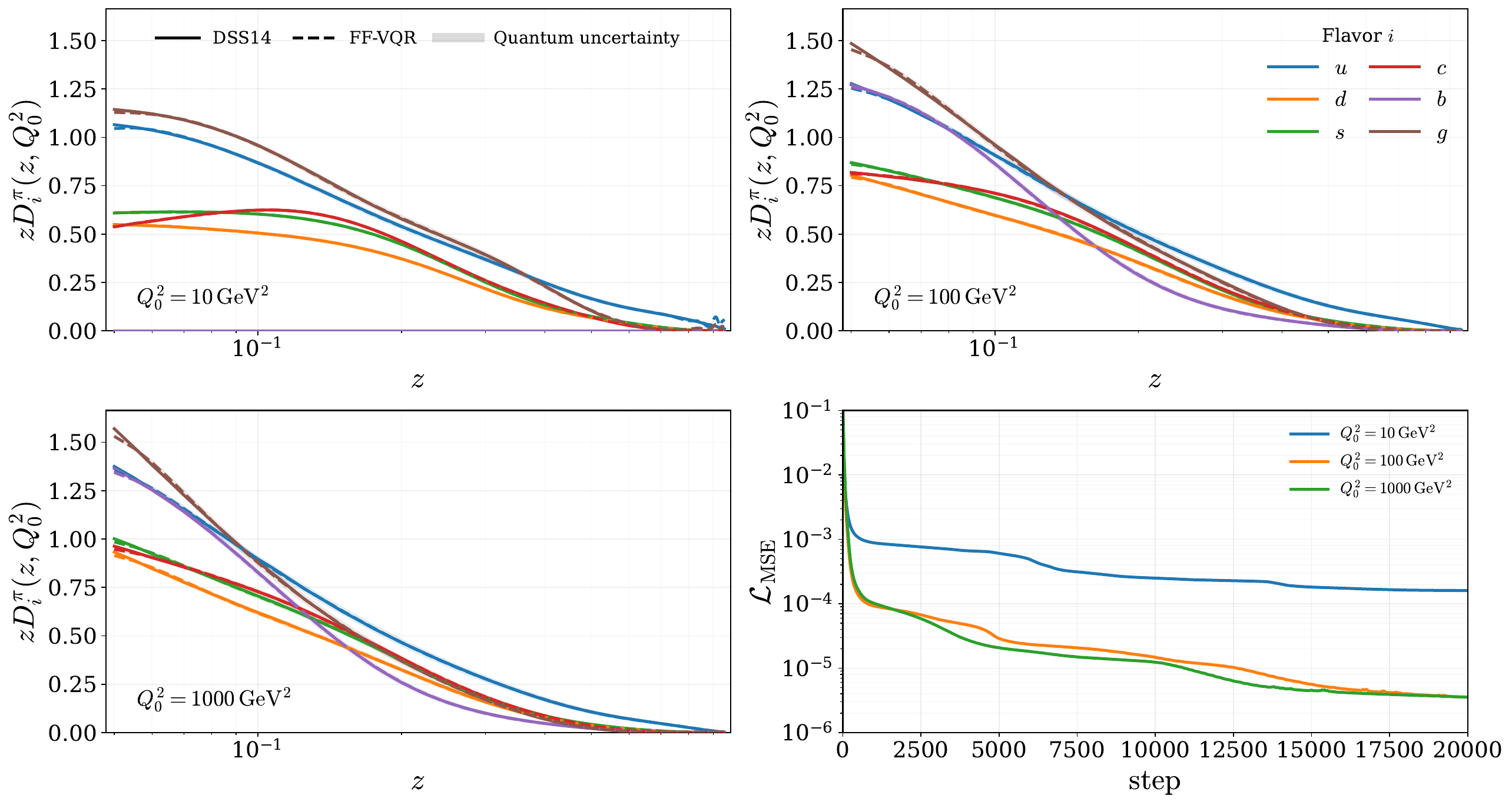}
    \caption{Comparison between DSS14 FFs and the corresponding FF-VQR at fixed scales $Q_0^2=10$, $100$, and $1000~\mathrm{GeV}^2$. The lower-right panel shows the evolution of the training loss function.}
    \label{fig:VQR1D}
\end{figure}
In Table~\ref{tab:vqr_depth_scan}, we summarize the reconstruction performance of the FF-VQR model for three representative energy scales, $Q_0^2=10$, $100$, and $1000~\mathrm{GeV}^2$, as a function of the number of layers in the VQC. For all scales considered, the FF-VQR achieves normalized MSEs below $10^{-3}$ using only a modest number of trainable parameters. A clear improvement is observed as the number of variational layers is increased. At $Q_0^2=10~\mathrm{GeV}^2$, the loss function decreases from $7.0\times10^{-4}$ with one layer to $3.0\times10^{-4}$ with two layers, corresponding to an improvement by a factor of approximately 2.3. A similar behavior is observed at $Q_0^2=100~\mathrm{GeV}^2$ and $Q_0^2=1000~\mathrm{GeV}^2$, where the reconstruction error is reduced by factors of about 13.8 and 7.2, respectively.

Increasing the quantum circuit depth beyond two layers produces moderate gains. For example, at $Q_0^2=10~\mathrm{GeV}^2$ the loss function decreases from $3.0\times10^{-4}$ to $1.6\times10^{-4}$ when the number of layers is increased from two to three (a reduction of approximately 47$\%$), while at $Q_0^2=1000~\mathrm{GeV}^2$ the improvement is from $5.8\times10^{-6}$ to $3.5\times10^{-6}$ (i.e. a reduction of 40$\%$). This behavior indicates that most of the relevant functional structure is already captured by relatively shallow quantum circuits, suggesting that the FF-VQR does not require large quantum circuits  to accurately reproduce the DSS14 interpolation grids. 

An additional trend can be observed across the different energy scales. The reconstruction error systematically decreases as $Q_0^2$ increases. For a fixed circuit depth, the smallest losses are obtained at $Q_0^2=1000~\mathrm{GeV}^2$, while the largest errors are found at $Q_0^2=10~\mathrm{GeV}^2$. This behavior is consistent with the fact that DGLAP evolution progressively smooths the FFs at higher scales, reducing local structures and making the distributions easier to approximate with a compact variational Ansatz. 

The reconstructed FFs are displayed in Fig.~\ref{fig:VQR1D}. Excellent agreement is observed between the DSS14 interpolation grids and the FF-VQR predictions for all flavors and energy scales considered. The variational model reproduces both the overall normalization and the detailed shape of the distributions throughout the full kinematic region. The lower-right panel shows the evolution of the training loss function during optimization. In all cases, the loss function decreases rapidly during the first few hundred optimization steps and subsequently converges towards a stable minimum, demonstrating a robust and efficient training procedure.

\section{Spectral analysis of FF-VQR at a fixed energy scale}
\label{sec:SpectralVQC}
The VQC framework introduced in the previous section shows an excellent performance reproducing DSS14 FFs. The reason behind this lies in the recipe used to build the VQC: with simple rules and a physics-inspired feature map, the methodology leads to a flexible and accurate model to describe FFs. In this section, we justify the expressivity of the proposed model by studying the spectral decomposition of FF-VQR and comparing its structure with the Ans\"atze used by classical FF analysis. On top of that, we will show that the same spectral analysis leads to approximations of FF-VQR that require a fraction of the parameters of the quantum model, as listed in Table \ref{tab:vqr_depth_scan}, which might ease the implementation of this framework in current quantum devices and simulators.

Historically, PDF and FF analyses \cite{HERAPDF20,NNPDF31, DSS07, deFlorian:2007ekg} have relied on flexible
parametrizations built upon the characteristic endpoint behavior,
\beq
 N_i\,z^{\alpha_i}(1-z)^{\beta_i}\, ,
 \label{eq:Euleriana}
\eeq
usually supplemented by polynomial factors. Although such parametrizations have been remarkably successful, their finite functional form may introduce a dependence on the particular Ansatz adopted at the initial scale. More recently, neural-network-based determinations have provided an alternative strategy specifically designed to relax these functional assumptions and reduce the parametrization bias. The encoding adopted in this work has been chosen strategically to retain this standard functional structure. In particular, the feature map introduced in Eq.~(\ref{eq:Representation}) naturally reproduces the same powers of \(z\) and \(1-z\) that appear in conventional PDF and FF fits. Since VQCs admit a partial Fourier representation, the accessible
frequency spectrum is determined by the data-encoding strategy,
whereas the remainder of the circuit architecture determines the
corresponding Fourier coefficients
~\cite{Schuld:2020enb,Wiedmann:2024fourier,
Atchade-Adelomou:2023mjf}. The trained FF-VQR can therefore be expanded as:
\begin{equation}
zD_i(z)
=
\langle0|
U^\dagger\!\left(\mathbf{x}(z),\bm{\theta}\right)
\hat O_i
U\!\left(\mathbf{x}(z),\bm{\theta}\right)
|0\rangle
=
\sum_{\bm{\omega}\in\Omega}
c_{i,\bm{\omega}}(\bm{\theta})
e^{\mathrm{i}\bm{\omega}\cdot\mathbf{x}(z)} ,
\end{equation}
where $\bm{\theta}$ denotes the set of trainable variational parameters $\{\alpha,\beta,\gamma,\phi\}$, $i$ labels the flavor channel, and $\mathrm{i}$ denotes the imaginary unit. The accessible frequency spectrum $\Omega\subset\mathbb{R}^n$ is determined by the data-encoding strategy, whereas the coefficients $c_{i,\bm{\omega}}$ depend on the remainder of the quantum circuit architecture and on the measurement observable. Then, using the definition of the feature vector introduced in Eq. (\ref{eq:Representation}), we can write
\begin{equation}
zD_i(z)
=
\sum_{\bm{\omega}\in\Omega}
\tilde c_{i,\bm{\omega}}(\bm{\theta})
e^{\mathrm{i}\omega_{1i}z}
z^{\,\mathrm{i}\omega_{2i}}
(1-z)^{\,\mathrm{i}\omega_{3i}},
\end{equation}
where the constant phase factor has been absorbed into the coefficients, i.e.
\[
\tilde c_{i,\bm{\omega}}(\bm{\theta})
=
c_{i,\bm{\omega}}(\bm{\theta})
e^{\mathrm{i}\omega_{4i}}.
\]
At this point, we can explicitly appreciate that FF-VQR consists of a superposition of functions like Eq. (\ref{eq:Euleriana}), which generalizes the parametrizations used by DSS14 and other FF fits. 

Let us go further and provide an analytical approximation of FF-VQR. For the sake of simplicity, we will consider FF-VQR with $L=3$ layers at a fixed energy scale $Q^2_0=10~\mathrm{GeV}^2$. In this case, we consider the reduced fundamental frequency set
$\Omega_{\rm fund}^{(i)}
=\{0,\pm\bm{\omega}_{i1},\pm\bm{\omega}_{i2},\pm\bm{\omega}_{i3}\}$
and write
\begin{eqnarray}
zD_i(z)
&=&
c_{i0}+
\sum_{\ell=1}^{3}
\left[
c_{i,\bm{\omega}_{i\ell}}
e^{\mathrm{i}\Theta_{i\ell}(z)}
+
c_{i,-\bm{\omega}_{i\ell}}
e^{-\mathrm{i}\Theta_{i\ell}(z)}
\right],
\end{eqnarray}
where $\Theta_{i\ell}(z)=\bm{\omega}_{i\ell}\cdot\mathbf{x}(z)$. Since the FFs are real, we have the constraint
$c_{i,-\bm{\omega}_{i\ell}}
=
c_{i,\bm{\omega}_{i\ell}}^{\,*}$.
Using Euler's formula, each conjugate pair is rewritten as a linear combination of cosine and sine functions,
\begin{equation}
c_{i,\bm{\omega}_{i\ell}}
e^{\mathrm{i}\Theta_{i\ell}}
+
c_{i,-\bm{\omega}_{i\ell}}
e^{-\mathrm{i}\Theta_{i\ell}}
=
A_{i\ell}\cos\Theta_{i\ell}
+
B_{i\ell}\sin\Theta_{i\ell},
\end{equation}
with
$A_{i\ell}=2\,\mathrm{Re}\!\left(c_{i,\bm{\omega}_{i\ell}}\right)$
and
$B_{i\ell}=-2\,\mathrm{Im}\!\left(c_{i,\bm{\omega}_{i\ell}}\right)$.
Therefore,
\begin{equation}
zD_i(z)
=
c_{i0}
+
\sum_{\ell=1}^{3}
\left[
A_{i\ell}\cos\Theta_{i\ell}(z)
+
B_{i\ell}\sin\Theta_{i\ell}(z)
\right].
\label{eq:SinCosExpansion}
\end{equation}
Finally, using
\begin{equation}
A_{i\ell}=R_{i\ell}\cos\delta_{i\ell},
\qquad
B_{i\ell}=R_{i\ell}\sin\delta_{i\ell},
\end{equation} 
the equivalent amplitude--phase representation is
\begin{equation}
zD_i(z)
=
c_{i0}+
\sum_{\ell=1}^{3}
R_{i\ell}
\cos\!\left[\Theta_{i\ell}(z)-\delta_{i\ell}\right],
\label{eq:RCosExpansion}
\end{equation}
where the amplitudes and phase shifts are given by
\begin{equation}
R_{i\ell}
=
\sqrt{A_{i\ell}^{\,2}+B_{i\ell}^{\,2}},
\qquad
\delta_{i\ell}
=
\operatorname{atan}\!\left(B_{i\ell}/A_{i\ell}\right).
\end{equation}

The phase functions $\Theta_{i\ell}(z)$ are completely determined by the optimized encoding parameters of the trained VQC and therefore remain fixed throughout the fitting procedure. Consequently, the only unknown quantities are the coefficients multiplying the trigonometric basis functions. Rather than fitting the nonlinear amplitude--phase representation directly, we first determine the coefficients in the linear basis, namely Eq.~(\ref{eq:SinCosExpansion}). For a given flavor, the values of the basis functions are evaluated on the sampling points $\{z_n\}$, defining the design matrix
\begin{equation}
\mathbf{M}
=
\begin{pmatrix}
1 &
\cos\Theta_{i1}(z_1) &
\sin\Theta_{i1}(z_1) &
\cdots &
\cos\Theta_{i3}(z_1) &
\sin\Theta_{i3}(z_1)
\\
\vdots & \vdots & \vdots &\ddots & \vdots & \vdots
\\
1 &
\cos\Theta_{i1}(z_N) &
\sin\Theta_{i1}(z_N) &
\cdots &
\cos\Theta_{i3}(z_N) &
\sin\Theta_{i3}(z_N)
\end{pmatrix},
\end{equation}
and the vector of unknown coefficients
\begin{equation}
\mathbf{c}
=
\left(
c_{i0},
A_{i1},
B_{i1},
\ldots,
A_{i3},
B_{i3}
\right)^T.
\end{equation}
The optimal coefficients are obtained from the linear least-squares problem:
\begin{equation}
\mathbf{M}\mathbf{c}
=
\mathbf{D},
\end{equation}
whose minimum-norm solution is computed through the Moore--Penrose pseudoinverse,
\begin{equation}
\mathbf{c}
=
\mathbf{M}^{+}\mathbf{D},\quad
\bm M^+=\bm Q_2\bm\Sigma^+\bm Q_1^T\,,
\end{equation}
where ${\bf{Q}}_{1,2}$ are orthogonal matrices, and $\bf{\Sigma}$ is a diagonal matrix consisting of singular values
of $\bf{M}$, and ${\bf{\Sigma}}^+$ is simply the diagonal matrix consisting of reciprocals of
the non-zero singular values of ${\bf{M}}$. Finally, the coefficients are rewritten in the equivalent amplitude--phase form, 
yielding the compact representation in Eq.~(\ref{eq:RCosExpansion}).

The linear least-squares procedure determines the coefficients of the sine--cosine basis, from which the equivalent amplitude--phase representation is obtained algebraically. Conversely, the complex Fourier coefficients associated with the Euler expansion are uniquely determined through
\begin{equation}
c_{i,\bm{\omega}_{i\ell}}
=
\frac{R_{i\ell}}{2}
e^{-\mathrm{i}\delta_{i\ell}},
\qquad
c_{i,-\bm{\omega}_{i\ell}}
=
\frac{R_{i\ell}}{2}
e^{\mathrm{i}\delta_{i\ell}},
\end{equation}
thereby
recovering the equivalent complex Fourier representation within the reduced spectral basis.

The spectral reconstruction described so far employs only the three fundamental encoding frequency vectors $\bm{\omega}_{i\ell}$ and their corresponding phase functions $\Theta_{i\ell}(z)$, with $\ell=1,2,3$. These frequency vectors are determined by the optimized encoding parameters of the trained VQC. This representation captures the dominant spectral content of the FFs and already provides an accurate approximation, as illustrated in Fig.~\ref{fig:mixed_modes_reconstruction}. 

\subsection{Mixed modes and entanglement}
\label{ssec:MixedModes}
Despite providing a very good approximation to FF-VQR, small deviations are still present. This motivates us to investigate additional frequency components of the Fourier representation beyond the fundamental components retained in the reduced spectral reconstruction.

According to the Fourier analysis of parametrized quantum circuits, the accessible frequency spectrum is determined by the data encoding, whereas the remainder of the circuit architecture determines the corresponding Fourier coefficients~\cite{Schuld:2020enb}. In multiqubit quantum neural networks, the Fourier decomposition contains linear combinations of the fundamental encoding frequencies, while entanglement affects the corresponding Fourier coefficients~\cite{Panadero:2024qnn}.

In FF-VQR, the encoding frequencies are flavor dependent. Motivated by the linear combinations appearing in the multiqubit Fourier decomposition~\cite{Panadero:2024qnn}, we consider same-layer combinations involving different flavor channels, which we refer to as \emph{mixed modes}. We define the corresponding flavor-mixed frequency vectors as
\begin{equation}
\bm{\omega}_{ij,\ell}^{(\pm)}
=
\bm{\omega}_{i\ell}
\pm
\bm{\omega}_{j\ell},
\qquad \ell=1,2,3,
\end{equation}
with the associated phase functions
\begin{equation}
\Theta_{ij,\ell}^{(\pm)}(z)
=
\Theta_{i\ell}(z)
\pm
\Theta_{j\ell}(z)
=
\left(
\bm{\omega}_{i\ell}
\pm
\bm{\omega}_{j\ell}
\right)\cdot
\bm{x}(z).
\end{equation}

\begin{figure}[t]
    \centering
    \includegraphics[width=0.99\linewidth]{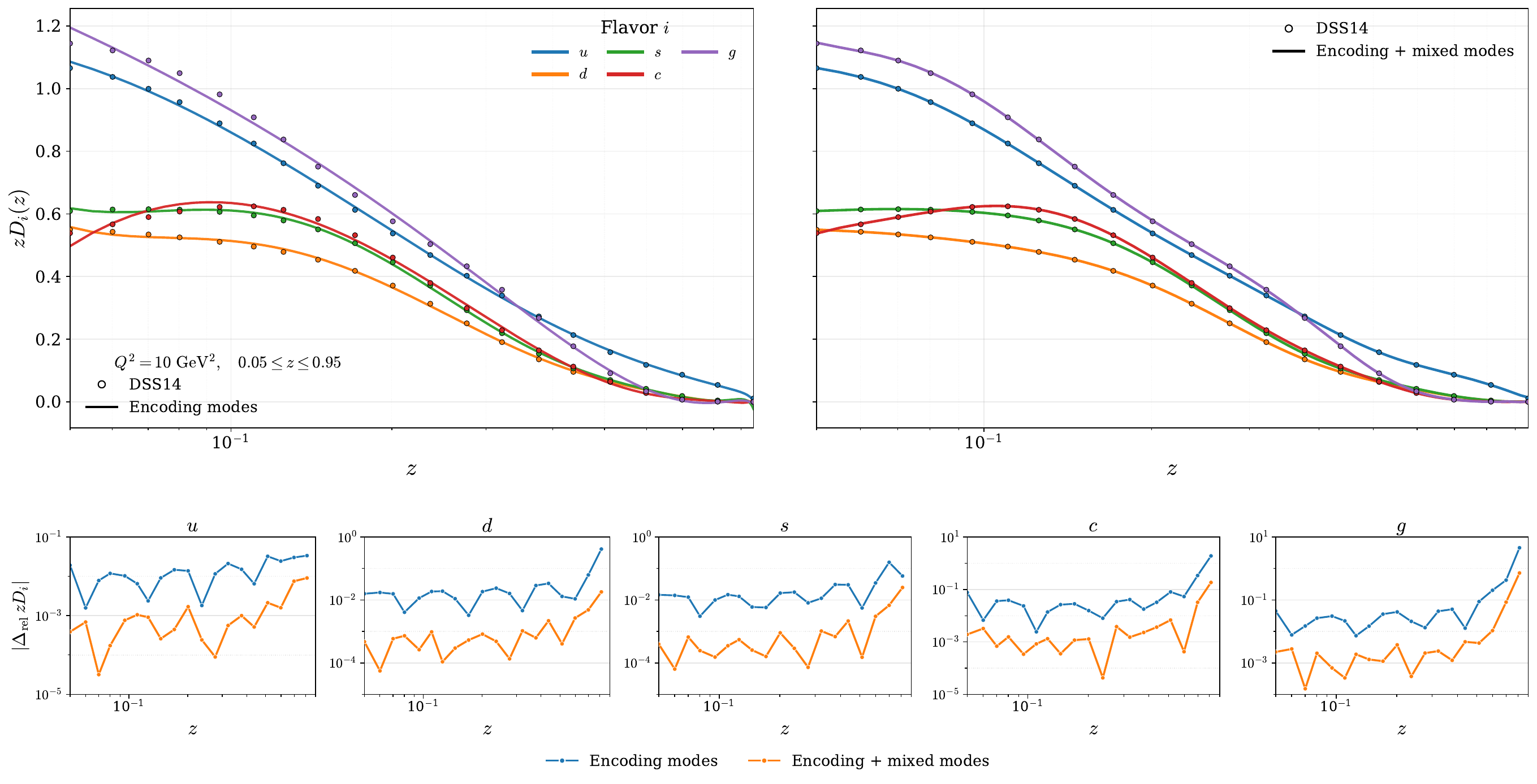}
    \caption{Comparison between the DSS fragmentation functions and their spectral
    reconstructions at $Q^2=10~\mathrm{GeV}^2$ for the five active flavor channels
    $u$, $d$, $s$, $c$, and $g$.
    In the upper-left panel, the DSS14 results (circles) are compared with the
    reconstruction obtained using only the fundamental encoding modes.
    The upper-right panel shows the corresponding reconstruction after adding
    the same-layer mixed modes considered for each flavor.
    The lower panels show the residuals
    $|\Delta_{\rm rel}\; zD_i(z)|$
    for each flavor. Blue curves correspond to the reconstruction using only the
    fundamental encoding modes, while orange curves include the mixed modes.
    The reduction of the residuals illustrates the contribution of the mixed
    modes to the spectral reconstruction.}
\label{fig:mixed_modes_reconstruction}
\end{figure}

The reduced Fourier expansion is therefore extended by
the corresponding same-layer mixed modes,
\begin{equation}
zD_i(z)
=
c_{i0}
+
\sum_{\ell=1}^{3}
R_{i\ell}
\cos\!\left[
\Theta_{i\ell}(z)-\delta_{i\ell}
\right]
+
\sum_{\ell=1}^{3}
R^{\rm mix}_{i\ell}
\cos\!\left[
\Theta^{\rm mix}_{i\ell}(z)-\delta^{\rm mix}_{i\ell}
\right],
\label{eq:mixed_series}
\end{equation}
where the additional amplitudes $R^{\rm mix}_{i\ell}$ and phase shifts
$\delta^{\rm mix}_{i\ell}$ are determined simultaneously with the
fundamental coefficients. For a quark flavor $i=u,d,s,c$, we consider
the same-layer difference with the gluon channel,
\begin{equation}
\bm{\omega}^{\rm mix}_{i\ell}
=
\bm{\omega}_{i\ell}
-
\bm{\omega}_{g\ell},
\qquad
\Theta^{\rm mix}_{i\ell}(z)
=
\Theta_{i\ell}(z)-\Theta_{g\ell}(z),
\qquad \ell=1,2,3.
\label{eq:mixed_quark_gluon}
\end{equation}
For the gluon channel, we analogously consider
\begin{equation}
\bm{\omega}^{\rm mix}_{g\ell}
=
\bm{\omega}_{g\ell}
-
\bm{\omega}_{u\ell},
\qquad
\Theta^{\rm mix}_{g\ell}(z)
=
\Theta_{g\ell}(z)-\Theta_{u\ell}(z),
\qquad \ell=1,2,3.
\label{eq:mixed_gluon_up}
\end{equation}
Thus, the index $j$ in the general mixed-frequency definition refers to a
different flavor channel, and the mixed modes used in the reconstruction
correspond to specific cross-channel combinations of the
flavor-dependent encoding frequencies.

To determine the contribution associated with these mixed frequencies,
we augment the linear basis by including the corresponding cosine and sine components,
\begin{equation}
\mathcal{B}_i
=
\Bigl\{
1,\,
\cos\Theta_{i\ell},\,
\sin\Theta_{i\ell},\,
\cos\Theta^{\rm mix}_{i\ell},\,
\sin\Theta^{\rm mix}_{i\ell}
\Bigr\}
_{\ell=1,2,3},
\end{equation}
and solve the corresponding least-squares problem through the
Moore--Penrose pseudoinverse (see App.~\ref{app:parameters}). As in the
fundamental reconstruction discussed above, the sine--cosine coefficients
are subsequently rewritten in the equivalent amplitude--phase form
$(R^{\rm mix}_{i\ell},\delta^{\rm mix}_{i\ell})$ appearing in
Eq.~(\ref{eq:mixed_series}).

The three same-layer mixed phase functions defined above are included simultaneously for each FF flavor. The resulting spectral reconstruction is shown in Fig.~\ref{fig:mixed_modes_reconstruction}. While the model containing only three fundamental phase functions already reproduces the DSS14 FFs with high accuracy, the inclusion of the mixed phase functions reduces significantly the remaining systematic deviations. The comparison therefore provides a direct estimate of the impact of the mixed spectral components with respect to the spectral reconstruction based only on the fundamental encoding modes. Notice that this analysis is performed at $Q^2=10~\mathrm{GeV}^2$, where the bottom-quark channel is inactive. The mixed-mode analysis is therefore performed for the five active flavor channels $i=\{u,d,s,c,g\}$.

To compare the results, we define the absolute relative differences as
\begin{equation}
|\Delta_{\rm rel}\, zD_i(z)|^{\rm enc}
=
\left|
\frac{
zD_i^{\mathrm{DSS14}}(z)
-
zD_i^{\mathrm{enc}}(z)
}{
zD_i^{\mathrm{DSS14}}(z)
}
\right|,
\end{equation}
and
\begin{equation}
|\Delta_{\rm rel}\, zD_i(z)|^{\rm mix}
=
\left|
\frac{
zD_i^{\mathrm{DSS14}}(z)
-
zD_i^{\mathrm{mix}}(z)
}{
zD_i^{\mathrm{DSS14}}(z)
}
\right|.
\end{equation}
Here, $zD_i^{\rm enc}(z)$ denotes the reconstructed FF using only the fundamental frequencies defined in Eq.~(\ref{eq:RCosExpansion}), while $zD_i^{\rm mix}(z)$ corresponds to the spectral reconstruction obtained after including the mixed modes defined in Eq.~(\ref{eq:mixed_series}).

These results show that a compact reduced Fourier representation provides an
accurate approximation to the trained VQC. While the three fundamental
encoding phase functions already provide an accurate reconstruction of the
DSS14 FFs, with validation coefficients $R^2$ ranging from $0.9985$ to
$0.9996$, the inclusion of the mixed modes systematically improves the
reconstruction to $R^2 \gtrsim 0.99999$ for all active flavor channels.
The relative $L_2$ error\footnote{This metric is defined in
App.~\ref{app:error}.} obtained using only the fundamental modes is
$E_{\rm enc}^{(u)}=5.33\times10^{-2}$ for the $u$ channel, while the
inclusion of the mixed modes reduces it to
$E_{\rm mix}^{(u)}=1.54\times10^{-2}$. This corresponds to an improvement
factor of
$F_u=E_{\rm enc}^{(u)}/E_{\rm mix}^{(u)}\simeq3.47$, or a reduction of
approximately $71.2\%$. For the $s$ channel, the relative $L_2$ error
decreases from $E_{\rm enc}^{(s)}=6.03\times10^{-2}$ to
$E_{\rm mix}^{(s)}=2.87\times10^{-2}$, corresponding to an improvement
factor of $F_s\simeq2.10$, or a reduction of approximately $52.4\%$.
Improvements are also observed for the remaining active flavors, with
relative-error reductions ranging from approximately $23.8\%$ for the
gluon channel to $73.4\%$ for the $d$ channel. This behavior is also
visible in Fig.~\ref{fig:mixed_modes_reconstruction}, where the absolute
relative differences are shifted to lower values over most of the $z$
range after including the mixed modes. Since the mixed phase functions
entering Eq.~(\ref{eq:mixed_series}) are constructed from the optimized
encoding frequencies of the trained VQC, the resulting expression provides
a compact analytical parametrization of the FFs at
$Q^2=10~\mathrm{GeV}^2$.

To conclude this section, we find that FF-VQR encoding can be efficiently approximated by exploiting its spectral structure. As discussed above, the accessible frequency spectrum $\Omega$ is determined by the data-encoding strategy~\cite{Schuld:2020enb}. For the present quantum architecture, with three encoding layers, the reduced spectral representation considered here contains six phase functions for each active flavor channel: three fundamental phase functions determined by the encoding frequencies and three same-layer mixed phase functions constructed from linear combinations of the fundamental frequencies.
For the quark channels, these mixed modes are associated with
$\Theta_{i\ell}-\Theta_{g\ell}$, whereas for the gluon channel they are given
by $\Theta_{g\ell}-\Theta_{u\ell}$, with $\ell=1,\ldots,3$.
This reduced spectral representation provides a compact description of
FF-VQR. Consequently, Eq.~(\ref{eq:mixed_series}), constructed from the
fundamental encoding frequencies and the mixed
phase functions
defined above, provides a compact approximate parametrization of the
FFs. Such a representation could serve as a
non-perturbative input for subsequent DGLAP evolution and for future
FF's analyses implemented on quantum devices.

\section{Two-dimensional variational quantum representation}
\label{sec:sevqc_interpolator}
In this section, we introduce a VQC designed to interpolate the DSS14 charged-pion FF's parametrization over the two-dimensional kinematic domain $(z,Q^2)$, extending the FF-VQR presented in the previous section. The purpose of this two-dimensional study is not to replace the DGLAP evolution, but to assess whether a VQC can faithfully reproduce the scale dependence already encoded in the DSS14 interpolation grid. The two-dimensional FF-VQR model combines physics-inspired encoding of features with trainable quantum correlations, in order to construct a compact quantum representation of the FFs across all partonic channels. The proposed quantum architecture is specifically tailored to incorporate the physical properties of timelike QCD evolution. It includes logarithmic kinematic encodings, smooth heavy-flavor threshold activations, a dedicated quantum register for the gluon sector, and trainable entanglement among the different partonic flavors. These ingredients introduce prior physical information directly into the quantum circuit, reducing the complexity of the learning task while increasing the expressive power of the variational Ansatz. The complete variational operator is denoted by
\begin{equation}
U_{\rm VQC}(z,Q^2;\theta) =U_{\rm ent}(\gamma) U_{\rm train}(\beta)U_{\rm enc}(z,Q^2;\alpha),
\label{eq:Encoding}
\end{equation}
where $z$ is the hadron momentum fraction, $Q^2$ is the hard scale, and $\theta\in\{\alpha,\beta,\gamma\}$ represents the full set of trainable variational parameters. The quantum circuit is designed to reproduce the DSS14 charged-pion FFs, i.e.
\begin{equation}
zD_i^{\pi^+}(z,Q^2),
\end{equation}
for the reduced flavor basis
\begin{equation}
i\in
\left\{
u,d,s,c,b,g
\right\}.
\end{equation}
The resulting quantum state is generated according to 
\begin{equation}
|\psi(z,Q^2;\theta)\rangle
=
U_{\rm VQC}(z,Q^2;\theta)
|0\rangle^{\otimes 6}.
\end{equation}
The VQC is organized into four building blocks, as illustrated in Fig.~\ref{fig:VQC2d}: (i) a physics-informed kinematic encoding layer, (ii) mass threshold encoding, (iii) rotation with free parameters, and  (iv) entanglement. The details of each component are presented in the following subsections.

\subsection{Variational quantum architecture}
\label{ssec:VQA}
\begin{figure}[!b]
    \centering
    \includegraphics[width=0.99\linewidth]{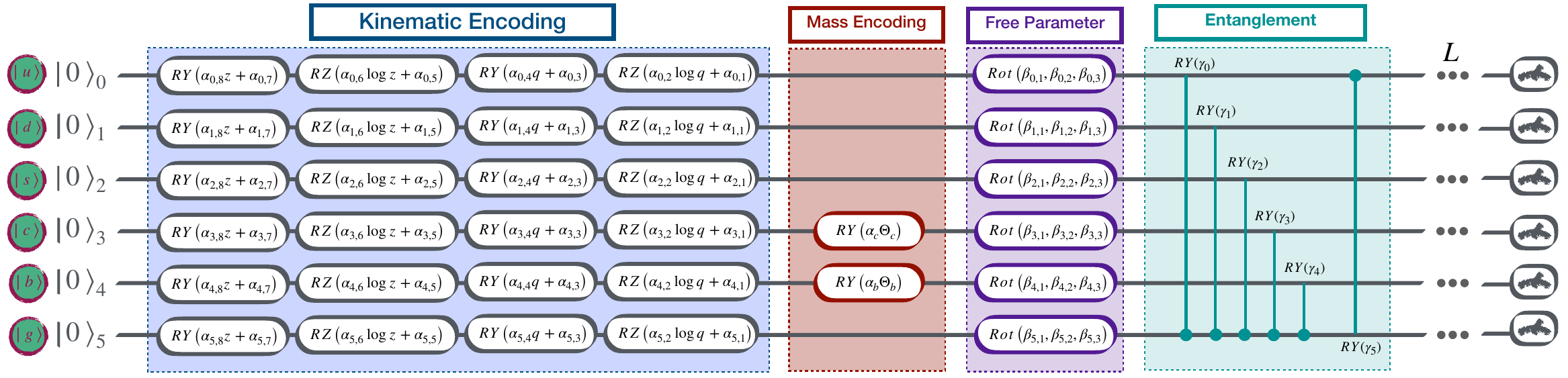}
    \caption{Pipeline and building-blocks of the two-dimensional FF-VQR.}
    \label{fig:VQC2d}
\end{figure}
The final VQC for the two-dimensional FF-VQR has been designed to combine problem-inspired input encoding with a trainable entanglement structure, allowing the quantum circuit to efficiently represent the multidimensional FF interpolator $z D_i^{\pi^+}(z,\,Q^2)$. The FFs for negatively charged and neutral pions are reconstructed using charge-conjugation symmetry and isospin relations, as explained in Sec.~\ref{sec:classical_determination}. The number of qubits is chosen to be $N_q=6$. This configuration was selected empirically, based on heuristic performance evaluations of this class of VQCs.

The proposed quantum architecture is shown in Fig.~\ref{fig:VQC2d}. The kinematic features $(z,\,q)$ are encoded into the quantum circuit, where $q$ is defined as the normalized variable
\beq
q=\frac{\ln{Q^2}-\ln {Q_{\min}^2}}{\ln{Q_{\max}^2}-\ln{Q_{\min}^2}}
\in[0,\,1].
\eeq
The encoding operator is defined as
\beq
\label{eq:encodif2d}
U_{i}(z,q)
=
R_Y(\alpha_{i8}z+\alpha_{i7})\,
R_Z(\alpha_{i6}\log z+\alpha_{i5})\,
R_Y(\alpha_{i4}q+\alpha_{i3})\,
R_Z(\alpha_{i2}\ln{q+\epsilon}+\alpha_{i1}),
\eeq
where $\epsilon=10^{-8}$ is introduced to avoid the logarithmic divergence at $q=0$ and $i$ labels the quark-flavor qubits following Eq.~(\ref{eq:FlavorLabel}). However, the heavy-quark FFs are generated only above their corresponding mass thresholds during the QCD evolution. These thresholds, located at $m_c=1.3$ GeV and $m_b=4.2$ GeV, produce a rapid change in the evolution of the corresponding FFs. To incorporate this threshold behavior into the quantum circuit, we introduce the sigmoid activation function:
\beq
\Theta_f(Q^2)
=
\sigma\left(\kappa\,\ln{Q^2/m_f^2}\right)
=
\frac{1}{1+(Q^2/m_f^2)^{-\kappa}}\,, \qquad f=b,c.
\label{eq:Sigmoid}
\eeq
This activation function requires an additional encoding gate,
\beq
R_Y(a_f\Theta_f),
\eeq
which injects information about the heavy-quark mass threshold into the quantum state. Further details are given in Sec.~\ref{ssec:heavyflavor}. This readout exploits both local expectation values and genuine two-qubit correlations without increasing the quantum circuit depth. The variational parameters are optimized by minimizing a MSE loss function defined over the reduced flavor basis. For a batch containing $N_{\rm batch}$ kinematic points, the loss function is given by
\beq
\mathcal L
=
\frac{1}{N_{\rm batch}}
\sum_{n=1}^{N_{\rm batch}}
\sum_{f}
\left(
\hat y_f^{(n)}
-
y_f^{(n)}
\right)^2,
\eeq
where $f=\{u,d,s,c,b,g\}$ labels the partonic flavors, $y_f$ denotes the normalized DSS14 target, and $\hat y_f\in[0,1]$ is the corresponding quantum prediction. 

\subsection{Quantum circuit implementation}

\label{ssec:implementation}
\begin{figure}[t]
    \centering
    \includegraphics[width=\linewidth]{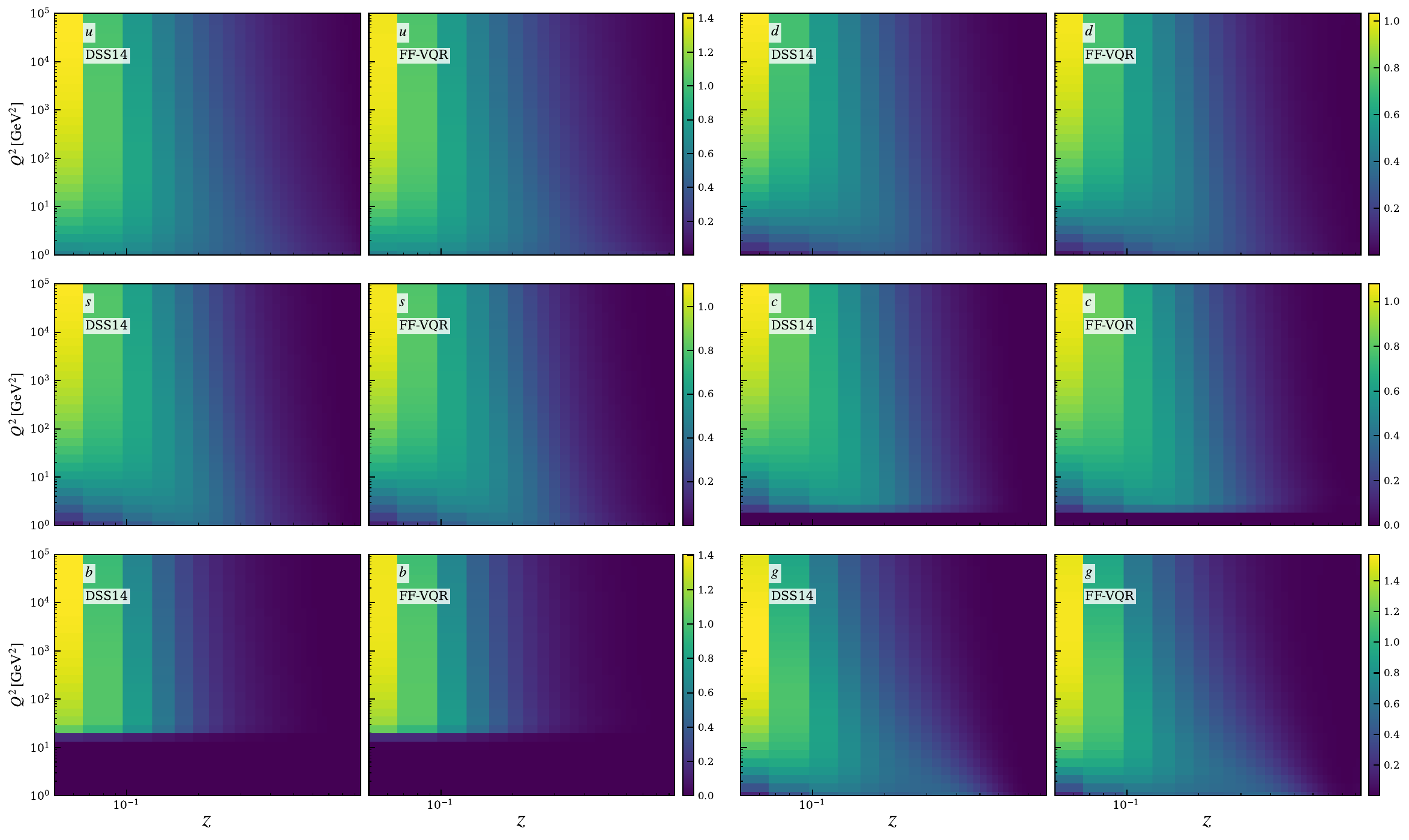}
    \caption{Comparison between DSS14 FFs and the corresponding FF-VQR of $z\, D_i^{\pi^+}(z,Q^2)$ in $(z,Q^2)$ space. From left to right, and top to bottom, we show $u$, $d$, $s$, $c$, $b$ and $g$-initiated FFs. We appreciate a rather good agreement between the two plots for each parton flavor. The thresholds of the heavy flavors are clearly visible.
  }
    \label{fig:VQR2D}
\end{figure}

\begin{figure*}[t]
    \centering
        \includegraphics[width=.7\linewidth
        ]{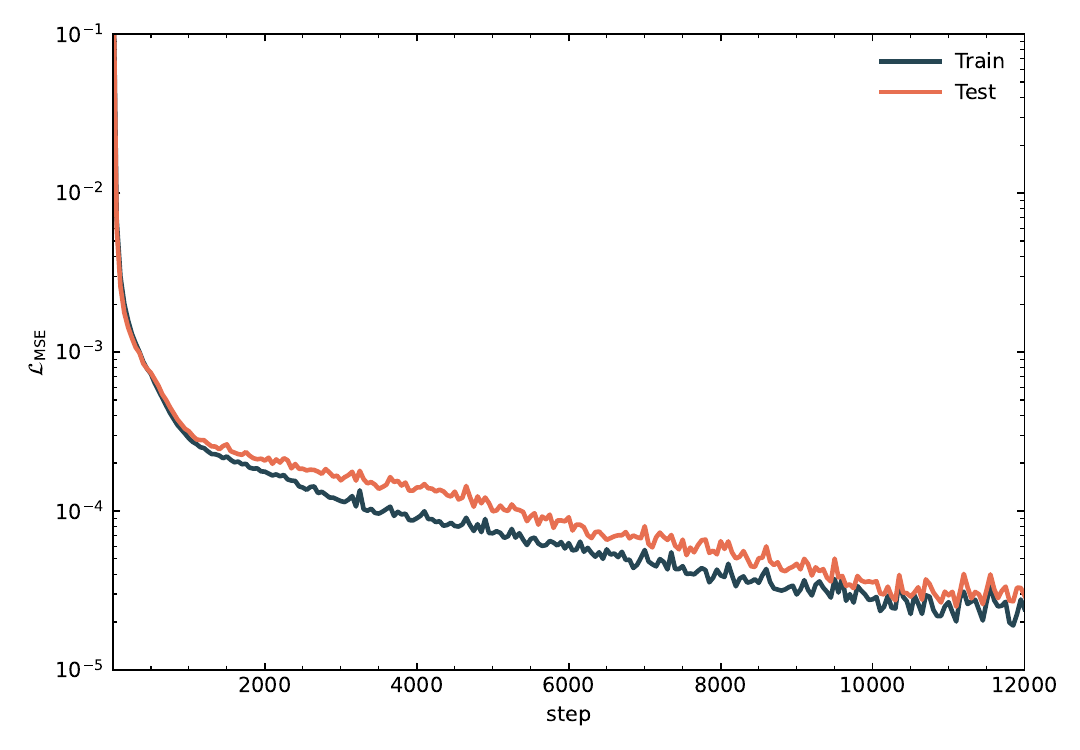}
    \caption{
    Evolution of the MSE loss function for the training and validation datasets during optimization of the two-dimensional FF-VQR.}
    \label{fig:sevqc_training_performance1}
\end{figure*}

The proposed quantum architecture is implemented using the
\texttt{default.qubit} statevector simulator provided by
\texttt{PennyLane}~\cite{Bergholm2018PennyLane}, while automatic differentiation and optimization are performed with \texttt{JAX}~\cite{jax2018github} in double precision. The DSS14 interpolation grid covers the kinematic domain $z\in[0.05,0.95]$ and  $Q^2\in[1,10^5]~{\rm GeV}^2$, using a uniform discretization of $N_z=30$, $N_{Q^2}=30$, resulting in a total of $N_{\rm dat}=900$ kinematic points. The dataset is randomly divided into $80\%$ training and $20\%$ validation samples. The optimization is performed using Adam~\cite{Kingma2015Adam} with ${\rm LR}=10^{-2}$, $N_{\rm batch}=256$ and $N_{\rm steps}=1.2\times10^{4}$, where each mini-batch is sampled randomly from the training set at every optimization step. The final implementation employs $L=6$ feature re-uploading layers connected through a ring of controlled-phase gates. The heavy-flavor threshold encoding is enabled for both charm and bottom quarks. The evolution of the MSE for both the training and validation samples is shown in Fig.~\ref{fig:sevqc_training_performance1}. The two curves
follow each other throughout the optimization, indicating stable
convergence and no evidence of overfitting. The final MSE reaches values of the order of $10^{-5}$ for both datasets, showing that the VQC successfully interpolates the DSS14
FFs over the complete range of $(z,Q^2)$. 

\begin{figure*}[t]
    \centering
        \includegraphics[width=\linewidth
        ]{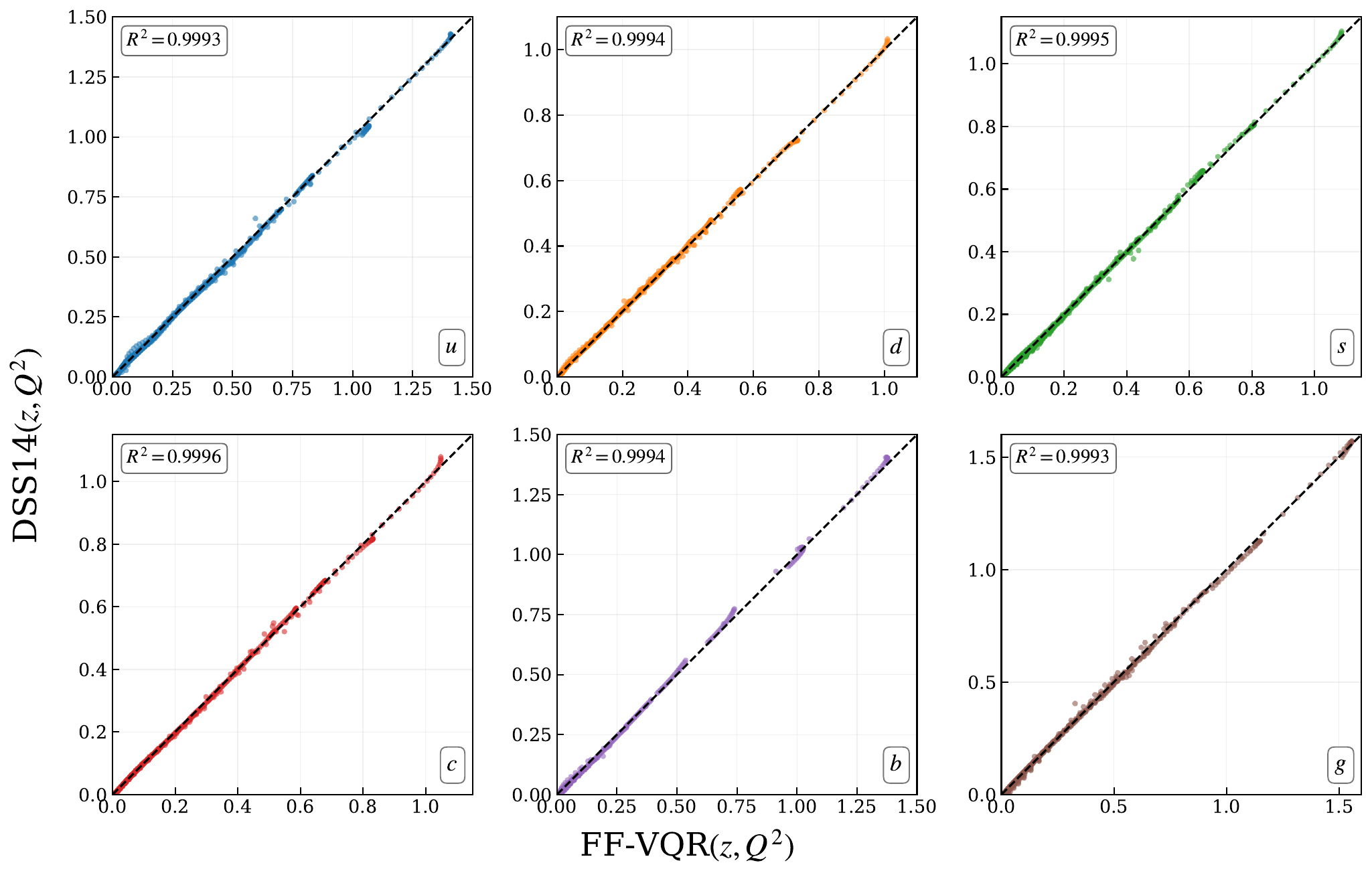}
    \caption{
    Performance of the two-dimensional FF-VQR: Point-by-point comparison between the DSS14 FFs of the $\pi^+$ and the corresponding FF-VQR predictions for the six independent partonic channels.}
    \label{fig:sevqc_training_performance2}
\end{figure*}

Finally, we study the correlations between the DSS14 FFs and the corresponding FF-VQR predictions. As depicted in Fig. \ref{fig:sevqc_training_performance2}, we perform a point-by-point comparison over the full $(z,Q^2)$ range used for the interpolation. For both quark- and gluon-initiated FFs, the regression coefficient is $R^2 > 0.999$. In all cases, this metric confirms the extremely high quality of the two-dimensional FF-VQR predictions.

\subsection{Heavy-flavor threshold encoding}
\label{ssec:heavyflavor}
The bottom and charm quark FFs exhibit a pronounced threshold at $Q^2\simeq m_b^2$ and $Q^2\simeq m_c^2$, respectively, where the corresponding heavy-quark channel becomes kinematically active. This localized transition is substantially sharper than the smooth scale dependence observed in the light-parton sector, making it one of the most challenging structures to reproduce with a shallow VQC. To incorporate this known QCD feature, the heavy-quark threshold function $\Theta_f(Q^2)$, defined in Eq.~(\ref{eq:Sigmoid}), is encoded into the quantum circuit through an additional $R_Y$ rotation. Rather than forcing the final prediction, this gate provides the variational Ansatz with explicit information about the location of the heavy-quark threshold. 

\begin{figure}[t]
\centering
\makebox[\textwidth][c]{%
    \includegraphics[height=4.5cm]{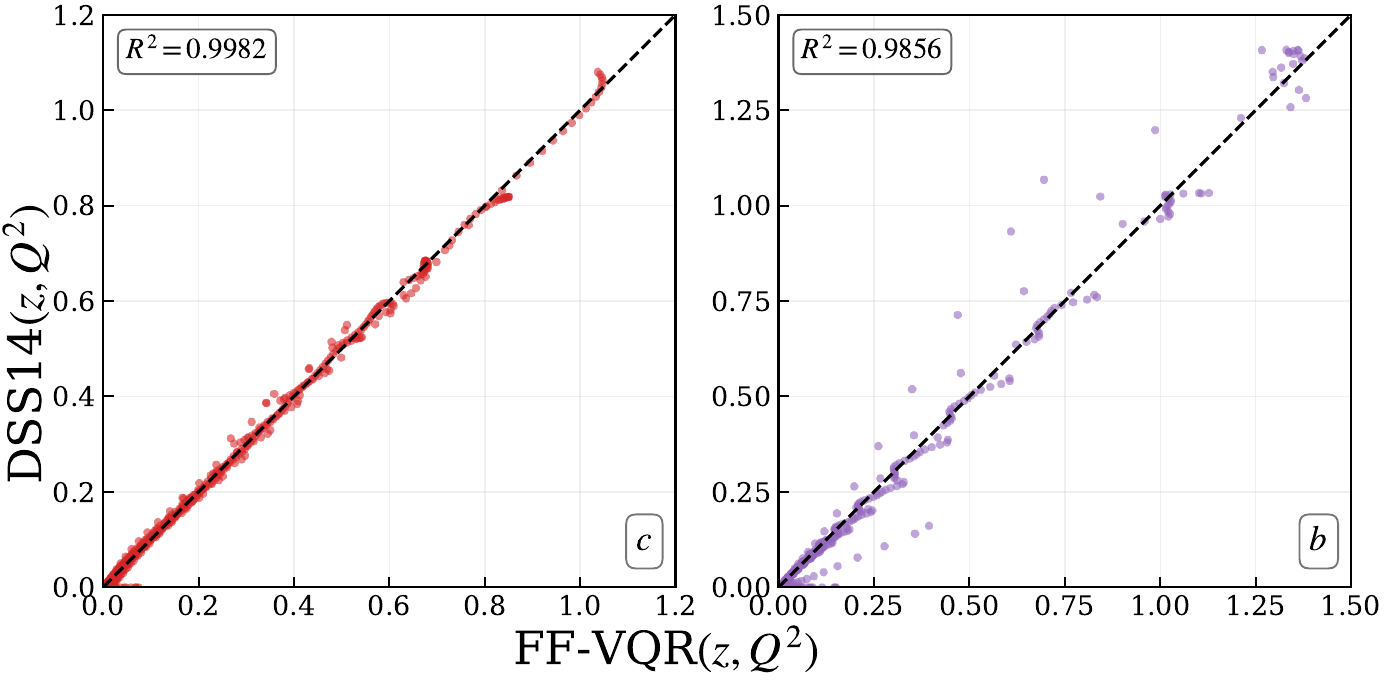}
    \includegraphics[height=4.3cm]{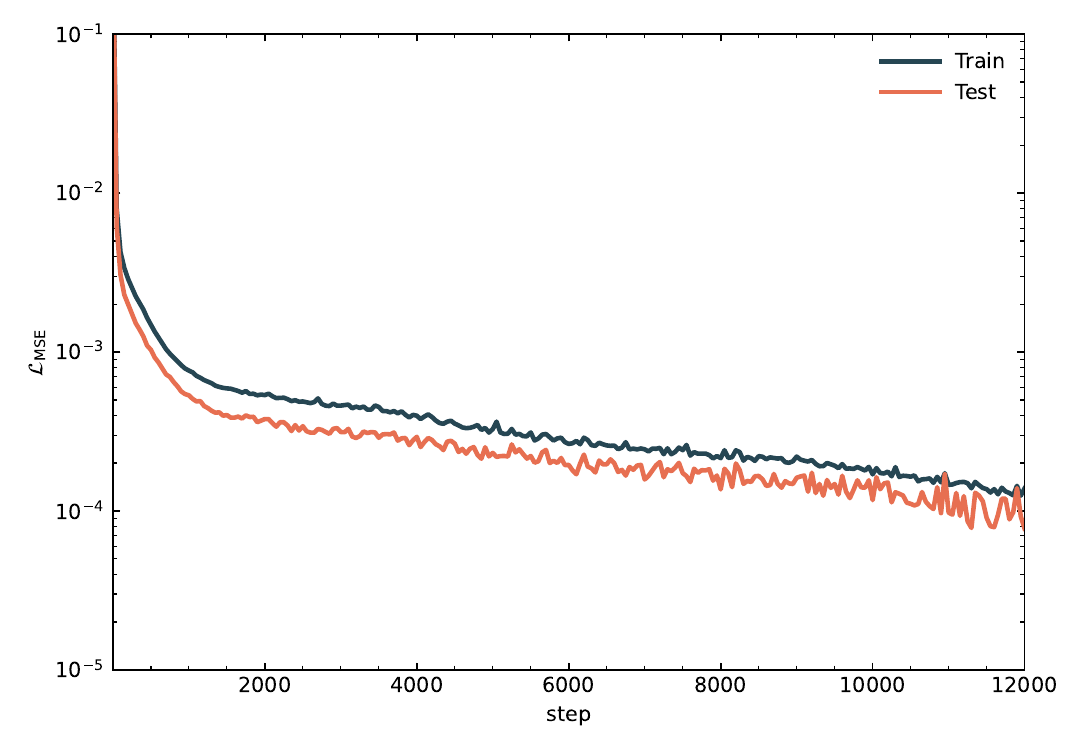}
}
\caption{
Effect of removing the heavy-flavor threshold encoding:
Correlation between the DSS14 bottom-quark FF and the corresponding FF-VQR prediction, and
training and validation weighted MSE loss function obtained using the same optimization setup after removing the bottom-quark threshold encoding.
}
\label{fig:bottom_corr}
\end{figure}

The importance of this encoding is assessed through an ablation study, where the threshold gate is removed while all remaining components of the model are kept unchanged, including the quantum circuit architecture, number of layers, optimizer, learning rate, mini-batch size, training dataset, initialization, and loss function. Consequently, any difference in the reconstruction can be attributed exclusively to the absence of the threshold encoding. The results are shown in Fig.~\ref{fig:bottom_corr}. 

We notice that removing the heavy-quark threshold encoding leads to a clear degradation in the interpolation quality. Although the global coefficient of determination remains high, the predictions develop a significantly larger dispersion around the ideal correlation line, particularly in the vicinity of the threshold where the FF changes most rapidly with $Q^2$. In this situation, the VQC is forced to infer the threshold location solely from the training data, instead of exploiting the known physical information encoded in the feature map. The training histories, shown in the right panel of Fig. \ref{fig:bottom_corr}, confirm that this behavior is not an optimization artifact. Since both models are trained under identical conditions, the observed deterioration originates entirely from removing the threshold encoding. This demonstrates that the heavy-quark activation acts as an effective physics-informed feature, reducing the complexity of the learning task while improving the reconstruction of the bottom-quark FF without increasing either the number of variational parameters or the quantum circuit depth.

\section{Conclusions}
\label{sec:Conclusions}
In this manuscript, we tackled the representation of fragmentation functions (FFs) through the design of optimized variational quantum circuits (VQCs). After a brief review of classical fitting methods, we encoded all the relevant information into parametric quantum circuits with physics-inspired Ansätze, including logarithmic kinematic encodings. Taking pion FFs as a case study, we exploited the physical symmetries of isospin and charge conjugation to identify an independent six-flavor basis of FFs that allowed us to describe $\pi^+$, $\pi^0$, and $\pi^-$ production. Moreover, we found that imposing these constraints was crucial for reducing redundancies in the VQCs and improving the convergence of the optimization.

The first implementation of such VQCs consisted of a one-dimensional representation of the FFs (FF-VQR) aimed at describing the $z$-dependence of the DSS14 set at a fixed energy scale. For this purpose, we introduced entanglement to connect gluon- and quark-initiated FFs, leading to much better agreement with the DSS14 predictions. Furthermore, we observed that the reconstruction quality was significantly improved using just two layers of parameters in the VQC. We also found that the MSE of the fit tends to decrease with increasing values of~$Q^2$.

In order to explain the excellent performance of the one-dimensional FF-VQR, we conducted a study of the spectral representation of the underlying quantum circuits. Relying on their Fourier representation, we found that the model is highly expressive while requiring access to only a few Fourier modes to reproduce DSS14 at a fixed energy scale. This provides further support for FF-VQR as a suitable and compact parametrization that could serve as a non-perturbative input for subsequent DGLAP evolution.

We then extended the FF-VQR model to reproduce the DSS14 charged-pion FFs over the two-dimensional kinematic domain $(z,Q^2)$. Our results show excellent agreement with DSS14 for $z \in [0.05,0.95]$ and $Q^2 \in [1,10^5] \ {\rm GeV}^2$, highlighting the expressive power of the two-dimensional FF-VQR model.

To conclude, the quantum architecture developed in this work may offer enhanced predictive power, precisely because the different flavor channels are intrinsically entangled within a common framework. Rather than introducing separate and decorrelated functional forms for each individual flavor, our quantum approach introduces a single quantum circuit to encode all the information required to describe the hadronization process. In this sense, the method may be viewed as providing an effective representation of the underlying hadronic structure, which is then projected onto the different possible partonic constituents. This viewpoint is conceptually novel and may open a new and efficient route for the description of both FFs and PDFs, while also facilitating their natural incorporation into the workflow of a quantum event generator.

\section*{Acknowledgments}
This work is supported by the Spanish Government and ERDF/EU - Agencia Estatal de Investigación MCIN/AEI/10.13039/501100011033, Grants No. PID2022-141910NB-I00, No. PID2023-146220NB-I00, No. EUR2025-164820, and No. CEX2023-001292-S. DFRE is supported by Generalitat Valenciana (CIGRIS/2022/145). RJHP is funded by SECIHTI through Project No.~CBF2023-2024-268 and Sistema Nacional de Investigadores. This work has benefited from activities carried out within the framework of the COST Action MLQC4FC (CA24146), supported by COST (European Cooperation in Science and Technology).

\appendix

\section{Parameters of the Spectral Fragmentation Model}
\label{app:parameters}
This appendix collects the numerical parameters entering the analytical reconstruction of the fragmentation functions introduced in Sec.~\ref{sec:SpectralVQC}. The Fourier model is entirely determined by two sets of quantities. First, the optimized encoding phases extracted from the trained VQC are written as
\begin{equation}
\Theta_i^{(\ell)}(z)
=
\omega_{1i}^{(\ell)}z
+
\omega_{2i}^{(\ell)}\log z
+
\omega_{3i}^{(\ell)}\log(1-z)
+
\omega_{4i}^{(\ell)},
\qquad
\ell=1,2,3,
\end{equation}
whose numerical coefficients are listed in Tab.~\ref{tab:encoding_coefficients}. These three fundamental phases constitute the basis from which the reduced Fourier representation is constructed. Using these phases together with the optimal mixed frequency identified by the spectral reduction procedure, the FFs are approximated by
\begin{equation}
zD_i(z)
=
c_{i0}
+
\sum_{\ell=1}^{3}
R_{i\ell}
\cos\!\left[
\Theta_{i\ell}(z)-\delta_{i\ell}
\right]
+
\sum_{\ell=1}^{3}
R^{\rm mix}_{i\ell}
\cos\!\left[
\Theta^{\rm mix}_{i\ell}(z)-\delta^{\rm mix}_{i\ell}
\right],
\label{eq:mixed_series2}
\end{equation}
where the corresponding amplitudes and phase shifts are summarized in Tables~\ref{tab:reduced_amplitudes} and \ref{tab:reduced_phases}. Together, Tables~\ref{tab:encoding_coefficients}, ~\ref{tab:reduced_amplitudes} and~\ref{tab:reduced_phases} provide all numerical information required to reconstruct the FF spectral decomposition at the reference scale $Q_0^2=10~\mathrm{GeV}^2$.

\begin{table*}[t]
\centering
\small
\setlength{\tabcolsep}{8pt}
\begin{tabular}{ccrrrr}
\midrule
Flavor & Mode &  $\omega_{1i}^{(\ell)}\,\,\,$  & $\omega_{2i}^{(\ell)}\,\,\,$ & $\omega_{3i}^{(\ell)}\,\,\,$ & $\omega_{4i}^{(\ell)}\,\,\,$ \\
\midrule
$u$ & $\Theta_{u}^{(1)}$ & $-$0.5730 & $-$0.4247 & 0.5437 & $-$0.4382 \\
$u$ & $\Theta_{u}^{(2)}$ & 0.5067 & $-$0.2002 & $-$0.2005 & $-$0.4864 \\
$u$ & $\Theta_{u}^{(3)}$ & $-$0.5142 & 0.1805 & 0.5580 & $-$0.5015 \\
$d$ & $\Theta_{d}^{(1)}$ & 0.4541 & 0.7545 & $-$0.1943 & 0.4170 \\
$d$ & $\Theta_{d}^{(2)}$ & 0.6632 & 0.5564 & 0.1949 & $-$0.2720 \\
$d$ & $\Theta_{d}^{(3)}$ & 0.2796 & 0.3891 & 0.3326 & 0.0690 \\
$s$ & $\Theta_{s}^{(1)}$ & 0.9233 & 1.0243 & $-$0.0279 & 0.7753 \\
$s$ & $\Theta_{s}^{(2)}$ & 1.1021 & 0.9820 & 0.0566 & $-$0.2852 \\
$s$ & $\Theta_{s}^{(3)}$ & 0.2361 & 0.8248 & $-$0.0921 & 0.0627 \\
$c$ & $\Theta_{c}^{(1)}$ & 1.2369 & 0.5175 & 0.3737 & 0.0937 \\
$c$ & $\Theta_{c}^{(2)}$ & 1.1914 & 0.4612 & $-$0.0959 & 0.0872 \\
$c$ & $\Theta_{c}^{(3)}$ & 0.4803 & 0.2819 & 0.2490 & 0.2338 \\
$g$ & $\Theta_{g}^{(1)}$ & 1.4464 & 0.2396 & $-$0.0151 & 0.2269 \\
$g$ & $\Theta_{g}^{(2)}$ & 1.9040 & 0.3741 & 0.3531 & $-$0.3965 \\
$g$ & $\Theta_{g}^{(3)}$ & 1.5640 & 0.0296 & 0.6179 & 0.0334 \\
\bottomrule
\end{tabular}
\caption{
Optimized coefficients defining the three fundamental encoding phases
$\Theta_i^{(\ell)}(z)
=
\omega_{1i}^{(\ell)}z
+
\omega_{2i}^{(\ell)}\log z
+
\omega_{3i}^{(\ell)}\log(1-z)
+
\omega_{4i}^{(\ell)}$
used in the reduced analytical reconstruction at $Q^2 = 10~\mathrm{GeV}^2$.
}
\label{tab:encoding_coefficients}
\end{table*}

\begin{table*}[t]
\centering

\scriptsize
\renewcommand{\arraystretch}{1.15}

\begin{tabular*}{\textwidth}{
@{\extracolsep{\fill}}
l
r
r r r
r r r
@{}
}
\toprule
Flavor
& $c_{i0}$
& $R_{i1}$
& $R_{i2}$
& $R_{i3}$
& $R^{\rm mix}_{i1}$
& $R^{\rm mix}_{i2}$
& $R^{\rm mix}_{i3}$
\\
\midrule

$u$
& $-382$
& $450$
& $674$
& $413$
& $126$
& $102$
& $792$
\\

$d$
& $1710$
& $583$
& $1560$
& $1150$
& $1010$
& $738$
& $2830$
\\

$s$
& $85.6$
& $514$
& $496$
& $762$
& $1120$
& $305$
& $36.5$
\\

$c$
& $-6690$
& $4080$
& $743$
& $14600$
& $8950$
& $4870$
& $12700$
\\

$g$
& $-463$
& $1570$
& $1230$
& $704$
& $353$
& $443$
& $1580$
\\

\bottomrule
\end{tabular*}

\caption{
Constant and amplitude coefficients of the reduced Fourier parametrization in \Eq{eq:mixed_series} at $Q_0^2=10\;{\rm GeV}^2$.
The coefficient $c_{i0}$ denotes the constant contribution,
$R_{i\ell}$ correspond to the three fundamental modes, and
$R^{\rm mix}_{i\ell}$ to the three same-layer mixed modes,
with $\ell=1,2,3$. }
\label{tab:reduced_amplitudes}
\end{table*}

\begin{table*}[t]
\centering
\scriptsize
\renewcommand{\arraystretch}{1.15}

\begin{tabular*}{\textwidth}{
@{\extracolsep{\fill}}
l
r r r
r r r
@{}
}
\toprule
Flavor
& $\delta_{i1}$
& $\delta_{i2}$
& $\delta_{i3}$
& $\delta^{\rm mix}_{i1}$
& $\delta^{\rm mix}_{i2}$
& $\delta^{\rm mix}_{i3}$
\\
\midrule

$u$
& $2.39$
& $-1.36$
& $-0.37$
& $0.41$
& $-0.85$
& $-2.60$
\\

$d$
& $-1.52$
& $2.43$
& $1.33$
& $2.65$
& $-2.61$
& $0.41$
\\

$s$
& $-2.99$
& $-0.60$
& $-2.52$
& $1.05$
& $-1.86$
& $1.72$
\\

$c$
& $1.05$
& $2.93$
& $-0.67$
& $-0.49$
& $2.71$
& $-2.87$
\\

$g$
& $-2.16$
& $0.12$
& $3.07$
& $-2.13$
& $0.93$
& $1.21$
\\

\bottomrule
\end{tabular*}

\caption{
Phase-shift coefficients of the reduced Fourier parametrization in \Eq{eq:mixed_series} at $Q_0^2=10\;{\rm GeV}^2$.
The parameters $\delta_{i\ell}$ correspond to the three fundamental modes, while $\delta^{\rm mix}_{i\ell}$ correspond to the three same-layer mixed modes, with $\ell=1,2,3$.
}
\label{tab:reduced_phases}
\end{table*}

\section{Accuracy metrics}
\label{app:error}
To characterize the accuracy of the spectral decomposition over the complete validation sample, we define the relative $L_2$ error as
\begin{equation}
E_{\rm rec}^{(i)}
=
\frac{
\sqrt{
\sum_n
\left[
zD_i^{\rm DSS14}(z_n)
-
zD_i^{\rm rec}(z_n)
\right]^2
}
}{
\sqrt{
\sum_n
\left[
zD_i^{\rm DSS14}(z_n)
\right]^2
}
},
\label{eq:relative_l2_error}
\end{equation}
where $zD_i^{\rm rec}(z)$ denotes either the spectral reconstruction obtained using only the fundamental encoding modes, $zD_i^{\rm enc}(z)$, or the spectral reconstruction including the mixed modes, $zD_i^{\rm mix}(z)$.

The improvement obtained after including the mixed modes is quantified by the factor
\begin{equation}
F_i
=
\frac{
E_{\rm enc}^{(i)}
}{
E_{\rm mix}^{(i)}
},
\label{eq:improvement_factor}
\end{equation}
where $E_{\rm enc}^{(i)}$ and $E_{\rm mix}^{(i)}$ are the relative $L_2$ errors of the fundamental and mixed-mode spectral reconstructions, respectively.

\bibliographystyle{JHEP}

\begin{thebibliography}{10}

\bibitem{Field:1976ve}
R.~D. Field and R.~P. Feynman, \emph{{Quark Elastic Scattering as a Source of High Transverse Momentum Mesons}}, \href{http://dx.doi.org/10.1103/PhysRevD.15.2590}{\emph{Phys. Rev. D} {\bf 15} (1977) 2590--2616}.

\bibitem{Collins:1989gx}
J.~C. Collins, D.~E. Soper and G.~F. Sterman, \emph{{Factorization of Hard Processes in QCD}}, \href{http://dx.doi.org/10.1142/9789814503266_0001}{\emph{Adv. Ser. Direct. High Energy Phys.} {\bf 5} (1989) 1--91}, [\href{http://arxiv.org/abs/hep-ph/0409313}{{\tt hep-ph/0409313}}].

\bibitem{Collins:2023cuo}
J.~Collins and T.~C. Rogers, \emph{{Definition of fragmentation functions and the violation of sum rules}}, \href{http://dx.doi.org/10.1103/PhysRevD.109.016006}{\emph{Phys. Rev. D} {\bf 109} (2024) 016006}, [\href{http://arxiv.org/abs/2309.03346}{{\tt 2309.03346}}].

\bibitem{ALEPH:1994cbg}
{\scshape ALEPH} collaboration, D.~Buskulic et~al., \emph{{Inclusive pi+-, K+- and (p, anti-p) differential cross-sections at the Z resonance}}, \href{http://dx.doi.org/10.1007/BF01556360}{\emph{Z. Phys. C} {\bf 66} (1995) 355--366}.

\bibitem{DELPHI:1998cgx}
{\scshape DELPHI} collaboration, P.~Abreu et~al., \emph{{pi+-, K+-, p and anti-p production in Z0 ---{\ensuremath{>}} q anti-q, Z0 ---{\ensuremath{>}} b anti-b, Z0 ---{\ensuremath{>}} u anti-u, d anti-d, s anti-s}}, \href{http://dx.doi.org/10.1007/s100529800989}{\emph{Eur. Phys. J. C} {\bf 5} (1998) 585--620}.

\bibitem{OPAL:1994zan}
{\scshape OPAL} collaboration, R.~Akers et~al., \emph{{Measurement of the production rates of charged hadrons in e+ e- annihilation at the Z0}}, \href{http://dx.doi.org/10.1007/BF01411010}{\emph{Z. Phys. C} {\bf 63} (1994) 181--196}.

\bibitem{SLD:1998coh}
{\scshape SLD} collaboration, K.~Abe et~al., \emph{{Production of pi+, K+, K0, K*0, phi, p and Lambda0 in hadronic Z0 decays}}, \href{http://dx.doi.org/10.1103/PhysRevD.59.052001}{\emph{Phys. Rev. D} {\bf 59} (1999) 052001}, [\href{http://arxiv.org/abs/hep-ex/9805029}{{\tt hep-ex/9805029}}].

\bibitem{TASSO:1988jma}
{\scshape TASSO} collaboration, W.~Braunschweig et~al., \emph{{Pion, Kaon and Proton Cross-sections in $e^+ e^-$ Annihilation at 34-{GeV} and 44-{GeV} Center-of-mass Energy}}, \href{http://dx.doi.org/10.1007/BF01555856}{\emph{Z. Phys. C} {\bf 42} (1989) 189}.

\bibitem{HERMES:2012uyd}
{\scshape HERMES} collaboration, A.~Airapetian et~al., \emph{{Multiplicities of charged pions and kaons from semi-inclusive deep-inelastic scattering by the proton and the deuteron}}, \href{http://dx.doi.org/10.1103/PhysRevD.87.074029}{\emph{Phys. Rev. D} {\bf 87} (2013) 074029}, [\href{http://arxiv.org/abs/1212.5407}{{\tt 1212.5407}}].

\bibitem{Makke:2013bya}
N.~Makke, \emph{{Fragmentation Functions measurement at COMPASS}}, \href{http://dx.doi.org/10.22323/1.191.0202}{\emph{PoS} {\bf DIS2013} (2013) 202}, [\href{http://arxiv.org/abs/1307.3407}{{\tt 1307.3407}}].

\bibitem{PHENIX:2003fvg}
{\scshape PHENIX} collaboration, S.~S. Adler et~al., \emph{{Mid-rapidity neutral pion production in proton proton collisions at $\sqrt{s}$ = 200-GeV}}, \href{http://dx.doi.org/10.1103/PhysRevLett.91.241803}{\emph{Phys. Rev. Lett.} {\bf 91} (2003) 241803}, [\href{http://arxiv.org/abs/hep-ex/0304038}{{\tt hep-ex/0304038}}].

\bibitem{BRAHMS:2007tyt}
{\scshape BRAHMS} collaboration, I.~Arsene et~al., \emph{{Production of mesons and baryons at high rapidity and high P(T) in proton-proton collisions at s**(1/2) = 200-GeV}}, \href{http://dx.doi.org/10.1103/PhysRevLett.98.252001}{\emph{Phys. Rev. Lett.} {\bf 98} (2007) 252001}, [\href{http://arxiv.org/abs/hep-ex/0701041}{{\tt hep-ex/0701041}}].

\bibitem{STAR:2006dgg}
{\scshape STAR} collaboration, J.~Adams et~al., \emph{{Forward neutral pion production in p+p and d+Au collisions at s(NN)**(1/2) = 200-GeV}}, \href{http://dx.doi.org/10.1103/PhysRevLett.97.152302}{\emph{Phys. Rev. Lett.} {\bf 97} (2006) 152302}, [\href{http://arxiv.org/abs/nucl-ex/0602011}{{\tt nucl-ex/0602011}}].

\bibitem{ALICE:2012wos}
{\scshape ALICE} collaboration, B.~Abelev et~al., \emph{{Neutral pion and $\eta$ meson production in proton-proton collisions at $\sqrt{s}=0.9$ TeV and $\sqrt{s}=7$ TeV}}, \href{http://dx.doi.org/10.1016/j.physletb.2012.09.015}{\emph{Phys. Lett. B} {\bf 717} (2012) 162--172}, [\href{http://arxiv.org/abs/1205.5724}{{\tt 1205.5724}}].

\bibitem{ALICE:2017nce}
{\scshape ALICE} collaboration, S.~Acharya et~al., \emph{{Production of ${\pi ^0}$ and $\eta $ mesons up to high transverse momentum in pp collisions at 2.76 TeV}}, \href{http://dx.doi.org/10.1140/epjc/s10052-017-4890-x}{\emph{Eur. Phys. J. C} {\bf 77} (2017) 339}, [\href{http://arxiv.org/abs/1702.00917}{{\tt 1702.00917}}].

\bibitem{ALICE:2021est}
{\scshape ALICE} collaboration, S.~Acharya et~al., \emph{{Nuclear modification factor of light neutral-meson spectra up to high transverse momentum in p\textendash{}Pb collisions at sNN=8.16 TeV}}, \href{http://dx.doi.org/10.1016/j.physletb.2022.136943}{\emph{Phys. Lett. B} {\bf 827} (2022) 136943}, [\href{http://arxiv.org/abs/2104.03116}{{\tt 2104.03116}}].

\bibitem{Borsa:2021ran}
I.~Borsa, D.~de~Florian, R.~Sassot and M.~Stratmann, \emph{{Pion fragmentation functions at high energy colliders}}, \href{http://dx.doi.org/10.1103/PhysRevD.105.L031502}{\emph{Phys. Rev. D} {\bf 105} (2022) L031502}, [\href{http://arxiv.org/abs/2110.14015}{{\tt 2110.14015}}].

\bibitem{deFlorian:2007ekg}
D.~de~Florian, R.~Sassot and M.~Stratmann, \emph{{Global analysis of fragmentation functions for protons and charged hadrons}}, \href{http://dx.doi.org/10.1103/PhysRevD.76.074033}{\emph{Phys. Rev. D} {\bf 76} (2007) 074033}, [\href{http://arxiv.org/abs/0707.1506}{{\tt 0707.1506}}].

\bibitem{deFlorian:2014xna}
D.~de~Florian, R.~Sassot, M.~Epele, R.~J. Hern\'andez-Pinto and M.~Stratmann, \emph{{Parton-to-Pion Fragmentation Reloaded}}, \href{http://dx.doi.org/10.1103/PhysRevD.91.014035}{\emph{Phys. Rev. D} {\bf 91} (2015) 014035}, [\href{http://arxiv.org/abs/1410.6027}{{\tt 1410.6027}}].

\bibitem{deFlorian:2017lwf}
D.~de~Florian, M.~Epele, R.~J. Hernandez-Pinto, R.~Sassot and M.~Stratmann, \emph{{Parton-to-Kaon Fragmentation Revisited}}, \href{http://dx.doi.org/10.1103/PhysRevD.95.094019}{\emph{Phys. Rev. D} {\bf 95} (2017) 094019}, [\href{http://arxiv.org/abs/1702.06353}{{\tt 1702.06353}}].

\bibitem{Borsa:2022vvp}
I.~Borsa, R.~Sassot, D.~de~Florian, M.~Stratmann and W.~Vogelsang, \emph{{Towards a Global QCD Analysis of Fragmentation Functions at Next-to-Next-to-Leading Order Accuracy}}, \href{http://dx.doi.org/10.1103/PhysRevLett.129.012002}{\emph{Phys. Rev. Lett.} {\bf 129} (2022) 012002}, [\href{http://arxiv.org/abs/2202.05060}{{\tt 2202.05060}}].

\bibitem{Bertone:2018ecm}
{\scshape NNPDF} collaboration, V.~Bertone, N.~P. Hartland, E.~R. Nocera, J.~Rojo and L.~Rottoli, \emph{{Charged hadron fragmentation functions from collider data}}, \href{http://dx.doi.org/10.1140/epjc/s10052-018-6130-4}{\emph{Eur. Phys. J. C} {\bf 78} (2018) 651}, [\href{http://arxiv.org/abs/1807.03310}{{\tt 1807.03310}}].

\bibitem{Moffat:2021dji}
{\scshape Jefferson Lab Angular Momentum (JAM)} collaboration, E.~Moffat, W.~Melnitchouk, T.~C. Rogers and N.~Sato, \emph{{Simultaneous Monte~Carlo analysis of parton densities and fragmentation functions}}, \href{http://dx.doi.org/10.1103/PhysRevD.104.016015}{\emph{Phys. Rev. D} {\bf 104} (2021) 016015}, [\href{http://arxiv.org/abs/2101.04664}{{\tt 2101.04664}}].

\bibitem{Soleymaninia:2020bsq}
M.~Soleymaninia, M.~Goharipour, H.~Khanpour and H.~Spiesberger, \emph{{Simultaneous extraction of fragmentation functions of light charged hadrons with mass corrections}}, \href{http://dx.doi.org/10.1103/PhysRevD.103.054045}{\emph{Phys. Rev. D} {\bf 103} (2021) 054045}, [\href{http://arxiv.org/abs/2008.05342}{{\tt 2008.05342}}].

\bibitem{Soleymaninia:2022qjf}
M.~Soleymaninia, H.~Hashamipour, H.~Khanpour and H.~Spiesberger, \emph{{Fragmentation functions for \ensuremath{\Xi}\ensuremath{-}/\ensuremath{\Xi}\textasciimacron{}+ using neural networks}}, \href{http://dx.doi.org/10.1016/j.nuclphysa.2022.122564}{\emph{Nucl. Phys. A} {\bf 1029} (2023) 122564}, [\href{http://arxiv.org/abs/2202.05586}{{\tt 2202.05586}}].

\bibitem{Soleymaninia:2024jam}
M.~Soleymaninia, H.~Hashamipour, M.~Salajegheh, H.~Khanpour, H.~Spiesberger and U.-G. Mei\ss{}ner, \emph{{Determination of $K^0_S$ Fragmentation Functions including BESIII Measurements and using Neural Networks}},  \href{http://arxiv.org/abs/2404.07334}{{\tt 2404.07334}}.

\bibitem{Soleymaninia:2026xjq}
{\scshape HAPS} collaboration, M.~Soleymaninia, H.~Khanpour, H.~Spiesberger, M.~Azizi, M.~Klasen and H.~Hashamipour, \emph{{Revisiting Unidentified Charged-Hadron Fragmentation Functions with Modern COMPASS SIDIS Multiplicities}},  \href{http://arxiv.org/abs/2605.31325}{{\tt 2605.31325}}.

\bibitem{Galvez-Viruet:2026jgx}
J.~J. G{\'a}lvez-Viruet, F.~J. Llanes-Estrada, N.~M. Arenaza, M.~G{\'o}mez-Rocha and T.~J. Hobbs, \emph{{First-principle predictions of fragmentation functions via quantum computing}},  \href{http://arxiv.org/abs/2608.30375}{{\tt 2608.30375}}.

\bibitem{Ball:2008by}
{\scshape NNPDF} collaboration, R.~D. Ball, L.~Del~Debbio, S.~Forte, A.~Guffanti, J.~I. Latorre, A.~Piccione et~al., \emph{{A Determination of parton distributions with faithful uncertainty estimation}}, \href{http://dx.doi.org/10.1016/j.nuclphysb.2008.09.037}{\emph{Nucl. Phys. B} {\bf 809} (2009) 1--63}, [\href{http://arxiv.org/abs/0808.1231}{{\tt 0808.1231}}].

\bibitem{Ball:2010de}
R.~D. Ball, L.~Del~Debbio, S.~Forte, A.~Guffanti, J.~I. Latorre, J.~Rojo et~al., \emph{{A first unbiased global NLO determination of parton distributions and their uncertainties}}, \href{http://dx.doi.org/10.1016/j.nuclphysb.2010.05.008}{\emph{Nucl. Phys. B} {\bf 838} (2010) 136--206}, [\href{http://arxiv.org/abs/1002.4407}{{\tt 1002.4407}}].

\bibitem{Carrazza:2019mzf}
S.~Carrazza and J.~Cruz-Martinez, \emph{{Towards a new generation of parton densities with deep learning models}}, \href{http://dx.doi.org/10.1140/epjc/s10052-019-7197-2}{\emph{Eur. Phys. J. C} {\bf 79} (2019) 676}, [\href{http://arxiv.org/abs/1907.05075}{{\tt 1907.05075}}].

\bibitem{NNPDF:2019vjt}
{\scshape NNPDF} collaboration, R.~Abdul~Khalek et~al., \emph{{A first determination of parton distributions with theoretical uncertainties}}, \href{http://dx.doi.org/10.1140/epjc/s10052-019-7364-5}{\emph{Eur. Phys. J.} {\bf C} (2019) 79:838}, [\href{http://arxiv.org/abs/1905.04311}{{\tt 1905.04311}}].

\bibitem{Perez-Salinas:2019pjx}
A.~P{\'e}rez-Salinas, A.~Cervera-Lierta, E.~Gil-Fuster and J.~I. Latorre, \emph{{Data re-uploading for a universal quantum classifier}}, \href{http://dx.doi.org/10.22331/q-2020-02-06-226}{\emph{Quantum} {\bf 4} (2020) 226}, [\href{http://arxiv.org/abs/1907.02085}{{\tt 1907.02085}}].

\bibitem{Perez-Salinas:2020nem}
A.~P\'erez-Salinas, J.~Cruz-Martinez, A.~A. Alhajri and S.~Carrazza, \emph{{Determining the proton content with a quantum computer}}, \href{http://dx.doi.org/10.1103/PhysRevD.103.034027}{\emph{Phys. Rev. D} {\bf 103} (2021) 034027}, [\href{http://arxiv.org/abs/2011.13934}{{\tt 2011.13934}}].

\bibitem{Schuld:2020enb}
M.~Schuld, R.~Sweke and J.~J. Meyer, \emph{{Effect of data encoding on the expressive power of variational quantum-machine-learning models}}, \href{http://dx.doi.org/10.1103/PhysRevA.103.032430}{\emph{Phys. Rev. A} {\bf 103} (2021) 032430}, [\href{http://arxiv.org/abs/2008.08605}{{\tt 2008.08605}}].

\bibitem{Cerezo2021VariationalQuantumAlgorithms}
M.~Cerezo, A.~Arrasmith, R.~Babbush, S.~C. Benjamin, S.~Endo, K.~Fujii et~al., \emph{Variational quantum algorithms}, \href{http://dx.doi.org/10.1038/s42254-021-00348-9}{\emph{Nature Reviews Physics} {\bf 3} (2021) 625--644}.

\bibitem{Ochoa-Oregon:2024zgm}
S.~A. Ochoa-Oregon, D.~F. Renter{\'\i}a-Estrada, R.~J. Hern{\'a}ndez-Pinto, G.~F.~R. Sborlini and P.~Zurita, \emph{{Using analytic models to describe effective PDFs}}, \href{http://dx.doi.org/10.1103/PhysRevD.110.036019}{\emph{Phys. Rev. D} {\bf 110} (2024) 036019}, [\href{http://arxiv.org/abs/2404.15175}{{\tt 2404.15175}}].

\bibitem{deLejarza:2025upd}
J.~J. Mart{\'\i}nez~de Lejarza, H.-Y. Wu, O.~Kyriienko, G.~Rodrigo and M.~Grossi, \emph{{Quantum Chebyshev probabilistic models for fragmentation functions}}, \href{http://dx.doi.org/10.1038/s42005-025-02361-1}{\emph{Commun. Phys.} {\bf 8} (2025) 448}, [\href{http://arxiv.org/abs/2503.16073}{{\tt 2503.16073}}].

\bibitem{Li:2024nod}
T.~Li, H.~Xing and D.-B. Zhang, \emph{{Simulating Parton Fragmentation on Quantum Computers}},  \href{http://arxiv.org/abs/2406.05683}{{\tt 2406.05683}}.

\bibitem{Wiedmann:2024fourier}
M.~Wiedmann, M.~Periyasamy and D.~D. Scherer, \emph{Fourier analysis of variational quantum circuits for supervised learning}, {\emph{arXiv preprint arXiv:2411.03450} (2024) }, [\href{http://arxiv.org/abs/2411.03450}{{\tt 2411.03450}}].

\bibitem{Atchade-Adelomou:2023mjf}
P.~Atchade-Adelomou and K.~Larson, \emph{{Fourier series weight in quantum machine learning}},  \href{http://arxiv.org/abs/2302.00105}{{\tt 2302.00105}}.

\bibitem{Dokshitzer:1977sg}
Y.~L. Dokshitzer, \emph{{Calculation of the Structure Functions for Deep Inelastic Scattering and e+ e- Annihilation by Perturbation Theory in Quantum Chromodynamics}}, {\emph{Sov. Phys. JETP} {\bf 46} (1977) 641--653}.

\bibitem{Gribov:1972ri}
V.~N. Gribov and L.~N. Lipatov, \emph{{Deep inelastic e p scattering in perturbation theory}}, {\emph{Sov. J. Nucl. Phys.} {\bf 15} (1972) 438--450}.

\bibitem{Altarelli:1977zs}
G.~Altarelli and G.~Parisi, \emph{{Asymptotic Freedom in Parton Language}}, \href{http://dx.doi.org/10.1016/0550-3213(77)90384-4}{\emph{Nucl. Phys. B} {\bf 126} (1977) 298--318}.

\bibitem{HERAPDF20}
H1 and Z.~Collaborations, \emph{Combination of measurements of inclusive deep inelastic $e^{\pm}p$ scattering cross sections and qcd analysis of hera data}, {\emph{Eur. Phys. J. C} {\bf 75} (2015) 580}.

\bibitem{NNPDF31}
R.~D.~B. et~al., \emph{Parton distributions from high-precision collider data}, {\emph{Eur. Phys. J. C} {\bf 77} (2017) 663}.

\bibitem{DSS07}
R.~S. D.~de Florian and M.~Stratmann, \emph{Global analysis of fragmentation functions for pions and kaons and their uncertainties}, {\emph{Phys. Rev. D} {\bf 75} (2007) 114010}.

\bibitem{Panadero:2024qnn}
I.~Panadero, Y.~Ban, H.~Espin{\'o}s, R.~Puebla, J.~Casanova and E.~Torrontegui, \emph{{Regressions on quantum neural networks at maximal expressivity}}, \href{http://dx.doi.org/10.1038/s41598-024-81436-5}{\emph{Sci. Rep.} {\bf 14} (2024) 31669}, [\href{http://arxiv.org/abs/2311.06090}{{\tt 2311.06090}}].

\bibitem{Bergholm2018PennyLane}
V.~Bergholm, J.~Izaac, M.~Schuld, C.~Gogolin, S.~Ahmed, V.~Ajith et~al., \emph{Pennylane: Automatic differentiation of hybrid quantum-classical computations},  \href{http://arxiv.org/abs/1811.04968}{{\tt 1811.04968}}.

\bibitem{jax2018github}
J.~Bradbury, R.~Frostig, P.~Hawkins, M.~J. Johnson, Y.~Katariya, C.~Leary et~al., \emph{{JAX}: composable transformations of {P}ython+{N}um{P}y programs},  2018.

\bibitem{Kingma2015Adam}
D.~P. Kingma and J.~Ba, \emph{Adam: A method for stochastic optimization},  in \emph{International Conference on Learning Representations (ICLR)}, 2015.
\newblock \href{http://arxiv.org/abs/1412.6980}{{\tt 1412.6980}}.

\end{thebibliography}
\providecommand{\href}[2]{#2}\begingroup\raggedright\endgroup

\end{document}